\documentstyle{mn}

\def\Aav{$\langle A \rangle$~}
\def\Pav{$\langle P_{ab} \rangle$~}
\title[Catalog of helium abundances]{A catalog of helium abundance
indicators from globular cluster photometry}

\author[E. L. Sandquist]{Eric L. Sandquist$^1$\thanks{Current address:
Department of Astronomy, San Diego State University, 5500 Campanile
Road, San Diego, CA~92182 USA; email: erics@mintaka.sdsu.edu}\\
$^1$Dearborn Observatory, Northwestern University, 2131 Sheridan Road,
Evanston, IL~60208 USA}

\date{}

\begin{document}

\maketitle

\begin{abstract}
We present a survey of helium abundance indicators derived from a
comprehensive study of globular cluster photometry in the literature.
For each of the three indicators used, we have conducted a thorough
error analysis, and identified systematic errors in the computational
procedures. For the population ratio $R = N_{HB} / N_{RGB}$, we find
that there is no evidence of a trend with metallicity, although there
appears to be real scatter in the values derived. Although this
indicator is the one best able to provide useful absolute helium
abundances, the mean value is $Y \approx 0.20$, indicating the
probable presence of additional systematic error.

For the magnitude difference from the horizontal branch to the main
sequence $\Delta$ and the RR Lyrae mass-luminosity exponent $A$, it is
only possible to reliably determine relative helium abundances. This
is due to continuing uncertainties in the absolute metallicity scale
for $\Delta$, and uncertainty in the RR Lyrae temperature scale for $A$. Both
indicators imply that the helium abundance is approximately constant
as a function of [Fe/H]. According to the $A$ indicator, both Oosterhoff I
and II group clusters have constant values independent of [Fe/H] and
horizontal branch type. In addition, the two groups have slopes $d \log
\mbox{\Pav} / d\mbox{[Fe/H]}$ that are consistent with each
other, but significantly smaller than the slope for the combined sample.
\end{abstract}


\section{Introduction}\label{intro}

The determination of helium abundances in stars has been a
long-standing problem in astronomy. For the old stars in globular
clusters and in the Galactic halo, it is only possible to
spectroscopically measure the helium abundance in a direct way for hot
horizontal branch (HB) stars. Studies of this kind (e.g. Moehler,
Heber, \& Durell 1997; Moehler, Heber, \& Rupprecht 1997) have found
abundances significantly higher {\it and} lower than the primordial
helium abundance determined from observations of low-metallicity
extragalactic H II regions (e.g. $Y_{P}= 0.234\pm0.002$ from Olive,
Skillman, \& Steigman 1997).  The interpretation typically given for the low
helium abundances is that gravitational settling has acted
preferentially on helium atoms in the atmospheres of these stars,
rendering the measurements unusable for determining the initial helium
abundance of the stars. High helium abundances may be the result of mass
loss before the HB phase, which could remove most of the hydrogen-rich
envelope of the stars.

Indirect methods must be used to measure the helium abundance in
earlier phases of stellar evolution. The most useful methods are applied
to globular clusters because individual clusters provide us with large
samples of stars with seemingly identical compositions and ages.
Globular cluster stars also seem to be among the oldest that can be found
in the Galaxy, so that if the helium abundance could be determined, it
would give us a hint of the primordial helium abundance.

There are two other reasons for examining the helium abundances of
globular clusters. First, it is important to look at the overall trend in
the helium abundance as a function of the cluster metal content. This
provides a check of our understanding of nucleosynthesis and Galactic
evolution. In addition, previous studies of RR Lyrae stars have indicated
that an anticorrelation of helium abundance with metal content can
explain the Sandage period-shift effect (Sandage 1982).

Second, it is important to see if it is possible to detect helium
variations among clusters of similar metallicity, since helium remains
a plausible candidate for the ``second parameter'' determining the
morphology of the HB. While age seems to play a role in some clusters,
it appears to be inadequate to explain all of the variations seen.
In the past, clusters have been marked as possibly having high helium
abundances (e.g., Dickens 1972), but to date there has not been
definitive evidence of a relative helium abundance difference.

We will examine three different helium abundance indicators that can
be measured from cluster photometry. We do this in the hopes that
anomalous clusters will appear with unusual values in two or more of
the parameters, so as to provide better evidence for variations. A
cursory examination of older tabulations of data reveals an additional
reason for recomputing values for these indicators: the trend of
helium abundance with metallicity appears to depend on the method used
(Caputo \& Castellani 1983).

The three indicators have been discussed to varying degrees in earlier
studies: $R$ (Iben 1968; Buzzoni et al 1983; Caputo, Martinez Roger,
\& Paez 1987), $A$ (Caputo, Cayrel, \& Cayrel de Strobel 1983;
hereafter, CCC), and $\Delta$ (Carney 1980; CCC).  Photometric studies
of globular clusters have become rather numerous in the past decade,
so that the size and quality of the dataset for each parameter can be
considerably increased. Where it has been possible, we have also
analyzed data for old globular clusters in the Magellanic Clouds, and
for stars in Local Group dwarf spheroidals with old populations.

In the following sections, we discuss each parameter in turn -- their
definitions, the datasets, potential errors in measurement, and
calibration. In the final two sections, we compare helium
abundance values derived from the different indicators, and examine
the evidence for trends as a function of [Fe/H].

\section{Helium Abundance Indicators}

For all three of the indicators discussed below, higher values imply
higher helium abundances.

\subsection{The Population Ratio $R$}\label{secr}

This indicator is simply defined as the ratio $R = N_{HB}/N_{RGB}$,
where $N_{HB}$ is the number of horizontal branch stars, and $N_{RGB}$
is the number of RGB stars brighter than the luminosity level of the
HB. This primarily reflects the dependence of the hydrogen-burning shell's
progress on the hydrogen content of the envelope material being fed into it
(and to a lesser extent, the change of helium core mass at helium
flash, which affects the HB luminosity, and hence, the HB lifetime).

Because the ratio is computed using the brightest stars in each
cluster, it can be calculated for any cluster having deep enough
photometry. However, because these stars are relatively scarce in
globular clusters, it usually requires wide-field data, or accurate
photometry of the cluster core. We have restricted our sample
to those clusters with photometric samples of at least a total of 100
HB and RGB stars. In addition, we require photometry of sufficient quality to
separate the HB, RGB, and AGB populations readily, and to
eliminate field stars from the sample when proper motion data is not
available.

We have chosen to use the ratio $R$ rather than the similar ratio
$R^{\prime} = N_{HB} / (N_{RGB} + N_{AGB})$, which would be more
easily measured in most cases, because $R$ is more straightforward to
calculate theoretically. For this reason, the stellar sample was
sometimes restricted in radius in order to remove portions of the
cluster center where image blending made identifications of HB, RGB,
and HB stars difficult.

In re-examining the determinations of $R$, we found that the vast
majority of earlier calculations used the ``average'' differential
bolometric correction $\Delta V_{BC} \equiv V_{RGB} - V_{HB}$ of 0.15
mag to determine the faint end of the RGB sample. This value
was used in the original Buzzoni et al. (1983) paper, but only because
uncertainties of various sorts made the correction unimportant.
However, the correction is actually a function of metallicity, and it
is applied to the faint end of the RGB sample, which means it can be
the source of significant systematic error.

To update $\Delta V_{BC}$, we have examined the HB models of Dorman
(1992) in conjunction with the isochrones of Bergbusch \& VandenBerg
(1992; hereafter BV92). These stellar models were computed with a
consistent set of physics and compositions. Although the composition
is somewhat out of date (it does not include full $\alpha$-element
enhancement), the differential bolometric corrections should be
satisfactory because they are differential in nature, and the oxygen
abundance does not affect the color of the giant branch relative to
the instability strip. We have determined the corrections as a
function of [Fe/H], and have derived the following fitting formula: \[
\Delta V_{BC} = 0.709 + 0.548 \mbox{[M/H]} +
0.229 \mbox{[M/H]}^{2} + 0.034 \mbox{[M/H]}^{3} .\] Because
$\alpha$-element enhancements influence the position of the HB and RGB
in the CMD like a change in [Fe/H] (Salaris, Chieffi, \& Straniero
1993), they must be taken into account when computing [M/H]. To do so,
we assumed a constant $\alpha$-element enhancement of 0.3 dex for all
of the globular clusters examined (Carney 1996). [The halo field
population is believed to show a different abundance pattern as a
function of metallicity (Wheeler, Sneden, \& Truran 1989).]  The
$\alpha$-element abundance was taken to contribute to the effective
metal content of the cluster as 0.7 [$\alpha$/Fe] (Salaris, Chieffi,
\& Straniero 1993). The contribution does not go directly as
[$\alpha$/Fe] because oxygen, the most abundant $\alpha$ element, has
a relatively high ionization potential. As a result, it does not
contribute as significantly to the opacity in the envelope of RGB
stars as it does for higher temperature HB stars.

Error in observed $R$ values comes from two sources: error in
the numbers of HB and RGB stars due to misidentification or
Poisson fluctuations, and error in the determination of the faint
limit for the RGB sample (resulting from the determination of the
magnitude level of the HB, or the metallicity uncertainty that
affects the size of the differential bolometric correction).

The magnitude level of the HB was taken to be the average magnitude of
stars in the instability strip. For clusters with many well-studied RR
Lyrae stars, this could be taken from the average magnitude of the
variables. Often though the RR Lyrae stars were not observed
frequently enough to derive good average magnitudes. If the cluster
had populated red and blue edges of the intability strip, a linear
interpolation between the edges was made. In cases where the cluster
had only a populated red or blue HB, a theoretical correction was made
using the models of Dorman (1992) to $\log_{10} T_{eff} = 3.85$ using
the cluster reddening and a well-determined point on the populated
portion of the HB. Dorman's scaled-solar abundance HB models were used 
with a correction for enhanced $\alpha$-element abundances as described above.

An examination of the table indicates that the quoted errors for the
HB magnitudes for the clusters are in general less than 0.05 mag. This
is because we are not interested in the {\it absolute} value of
$V_{HB}$ for any of the clusters -- only the value for the sample
used. It is clearly unnecessary for the photometric calibration of the
sample to be perfect, since we only require that the relative
photometry of the HB and RGB be good enough that the faint limit of
the RGB sample can be determined well. For this reason, the tabulated
values of $V_{HB}$ should not be taken as being good absolute
values. This column is provided to make verification of the results
easier.

The error due to the differential bolometric correction is
illustrative, so we present a semi-analytic model for its contribution
to the error. The important quantity to compute is $\partial N_{RGB} /
\partial V_{RGB}$. This can be derived assuming that i) the
differential luminosity function (LF) of the RGB has a constant slope
$\alpha = d(\log_{10}N)/dV$, and ii) the number of red giant branch
stars goes to zero at the tip of the RGB (TRGB). So, we use
$\log_{10}(N) = \alpha (V - V_{TRGB})$, and find \[ N_{RGB} =
\frac{N_{o}}{\ln 10} (10^{\alpha(V_{RGB}-V_{TRGB})} - 1) ,\] where
$N_{o}$ is the normalization of the luminosity function. From this we
find \[ \frac{\partial N_{RGB}}{\partial V_{RGB}} = \alpha \ln 10
\left[
\frac{10^{\alpha(V_{RGB}-V_{TRGB})}}{10^{\alpha(V_{RGB}-V_{TRGB})} -
1} \right] N_{RGB} .\] The fraction in brackets is close to one, and
from Bergbusch \& VandenBerg (1992) LFs we find $\alpha = 0.33$ nearly
independent of composition. From this we find that for each tenth of a
magnitude added to the faint limit of the RGB sample, the number of
RGB stars increases by about 9\%.  (Alternately, making the faint
limit fainter by one magnitude for a given cluster area increases the
total number of red giants by approximately 85\%.) This demonstrates
the importance of accurate values for the differential bolometric correction
$\Delta V_{BC}$.

The above derivative affects the error in $N_{RGB}$ through errors in
metallicity and in the determination of the HB magnitude. Because of the
metallicity dependence of the differential bolometric correction,
helium determinations for high metallicity clusters will be relatively
more uncertain. A polynomial fit to the derivative gives
\[ \frac{\partial \ln N_{RGB}}{\partial V_{RGB}} = 0.9469 +
0.1084\mbox{[M/H]} + 0.0257\mbox{[M/H]}^{2} .\] We note that the
error from this source goes as $N_{RGB}$, whereas the Poisson counting
errors go as $N_{RGB}^{1/2}$.

We have used [Fe/H] values from the tabulation of Djorgovski (1993) in
computing the values of $\Delta V_{BC}$ for the clusters. Djorgovski's
values fairly closely follow the values of Zinn \& West (1984;
hereafter, ZW). We have also considered an alternative metallicity scale
since this affects the differential bolometric
corrections. Using metallicities from Carretta \& Gratton (1996), we
recomputed the $R$ values for 16 clusters from their sample. The
results are shown in Fig.~\ref{ratcomp}. For the most part, the values
were little affected by the changes in the scale.

\begin{figure}
\vspace{5.5cm}
\caption{A comparison of $R$ values derived assuming metallicities
from the scale of Zinn \& West (1984; ZW) and Carretta \& Gratton
(1996; CG).}
\label{ratcomp}
\end{figure}

Computed $R$ values are insensitive to factors that merely shift the
RGB in temperature or color (age, for example). Factors that cause
changes in the absolute brightness of the HB (such as mean mass of the
helium cores of stars, and CNO abundances) will cause systematic
errors in measured values for individual clusters.

Our sample is composed of 42 Galactic globular clusters (GGCs) and 5
globulars in the Large Magellanic Cloud (LMC), and is presented in
Table~\ref{tratio}. The listings are grouped in bins of approximately
0.2 in [Fe/H]. References carrying a ``(PM)'' designation were used to
remove field stars from the sample using proper motion information.
There tend to be few clusters at the very metal-rich end due to a
combination of substantial field star contamination and confusion
between RGB and red HB cluster stars.

\begin{table*}
\begin{minipage}{15cm}
\caption{Data for the bright-star population ratio $R$}
\label{tratio}
\begin{tabular}{@{}lclcccll@{}}
ID (NGC/IC) & $N_{HB}$ & $V_{HB}$ & $\Delta V_{BC}$ & $N_{RGB}$ & $N_{AGB}$ &
$R$ & References \\
\hline
\multicolumn{8}{c}{\underline{Milky Way Clusters}}\\
5927 & 134 & $16.57\pm0.02^{a}$ & 0.66 & 168 & 20 &
$0.80\pm0.12$ & Samus et al 1996, \\
& & & & & & & Sarajedini \& Norris 1994 \\
& & & & & & & \\
6496 & 69 & $16.45\pm0.02^{a}$ & 0.62 & 53 & 5 &
$1.30\pm0.25$ & Richtler et al 1994 \\
6624 & 126 & $16.22\pm0.02^{a}$ & 0.63 & 177 & 30 &
$0.71\pm0.11$ & Richtler et al 1994 \\
6637 (M69) & 127 & $15.95\pm0.02^{a}$ & 0.53 & 101 &
21 & $1.26\pm0.19$ & Sarajedini \& Norris 1994\\
& & & & & & & \\
104 (47 Tuc) & 368 & $13.99\pm0.02^{a}$ & 0.49 & 307 & 38 &
$1.19\pm0.12$ & Lee 1977a \\
& & & & & & & \\
6171 (M107) & 117 & $15.62\pm0.02$ & 0.41 & 135 & 29 &
$0.87\pm0.12$ & Ferraro et al 1991, \\
& & & & & & & Dickens \& Rolland 1972, \\
& & & & & & & Cudworth et al. 1992 (PM)\\
6652 & 75 & $15.86\pm0.02^{a}$ & 0.41 & 61 & 20 &
$1.23\pm0.22$ & Ortolani et al 1994 \\
& & & & & & & \\
6712 & 122 & $16.30\pm0.02$ & 0.41 & 102 & 19 & $1.20\pm0.17$ &
Cudworth 1988 (PM)\\
6723 & 101 & $15.46\pm0.02$ & 0.38 & 79 & 15 & $1.28\pm0.20$ & Menzies 1974\\
& & & & & & & \\
362 & 94 & $15.42\pm0.02$ & 0.35 & 83 & 14 & $1.13\pm0.17$ & Harris 1982 \\
1261 & 148 & $16.64\pm0.03$ & 0.34 & 115 & 26 & $1.29\pm0.17$ & Ferraro et al 1993 \\
1851 & 209 & $16.15\pm0.02$ & 0.33 & 174 & 24 & $1.20\pm0.13$ & Walker 1992a \\
2808 & 247 & $16.19\pm0.03$ & 0.33 & 249 & 22 & $0.99\pm0.10$ & Ferraro et al 1990 \\
6121 (M4) & 114 & $13.39\pm0.03$ & 0.34 & 112 & 24 & $1.04\pm0.14$ &
Cudworth \& Rees 1990 (PM)\\
6266 (M62) & 114 & $16.40\pm0.03$ & 0.34 & 80 & 18 & $1.43\pm0.22$ &
Caloi et al 1987 \\
& & & & & & & \\
288 & 100 & $15.41\pm0.03^{a}$ & 0.33 & 85 & 20 & $1.18\pm0.18$ &
Buonanno et al 1984b \\
4499 & 145 & $17.65\pm0.04$ & 0.31 & 112 & 23 & $1.29\pm0.17$ &
Ferraro et al 1995 \\
5904 (M5) & 555 & $15.09\pm0.02$ & 0.33 & 515 & 94 & $1.08\pm0.07$
& Sandquist et al 1996 \\
& & & & & & & Cudworth 1979 (PM)\\
& & & & & & & Rees 1993 (PM)\\
6229 & 92 & $18.08\pm0.02$ & 0.31 & 100 & 19 & $0.92\pm0.14$ &
Carney et al 1991 \\ 
6681 (M70) & 82 & $15.61\pm0.02$ & 0.31 & 50 & 8 & $1.64\pm0.30$ &
Brocato et al 1996b \\
6752 & 225 & $13.70\pm0.04$ & 0.31 & 144 & 13 & $1.56\pm0.18$ &
Buonanno et al 1986 \\
6934 & 55 & $16.92\pm0.02$ & 0.31 & 62 & 12 & $0.89\pm0.17$ &
Brocato et al 1996b \\
6981 (M72) & 45 & $16.90\pm0.03$ & 0.31 & 56 & 10 & $0.80\pm0.16$ &
Brocato et al 1996b \\
7006 & 96 & $18.81\pm0.02$ & 0.30 & 62 & 24 & $1.55\pm0.26$ & Buonanno et al 1991\\
& & & & & & & \\
1904 (M79) & 122 & $16.20\pm0.03$ & 0.29 & 114 & 16 &
$1.07\pm0.14$ & Ferraro et al 1992b \\
3201 & 179 & $14.77\pm0.03$ & 0.30 & 150 & 14 & $1.19\pm0.14$ &
Lee 1977b \\
5272 (M3) & 562 & $15.66\pm0.03$ & 0.29 & 473 & 65 & $1.19\pm0.08$ &
Ferraro et al 1997 \\ 
5897 & 68 & $16.34\pm0.05^{a}$ & 0.29 & 50 & 8 & $1.36\pm0.26$ &
Ferraro et al 1992a \\
6093 (M80) & 170 & $15.51\pm0.05^{a}$ & 0.30 & 174 & 39 & $0.98\pm0.11$ & Alcaino et al 1998 \\
6205 (M13) & 90 & $14.89\pm0.03$ & 0.29 & 88 & 12 & $1.02\pm0.16$ & Cohen et al 1997 \\
6218 (M12) & 91 & $14.58\pm0.03^{a}$ & 0.30 & 56 & 12 & $1.63\pm0.28$
& Brocato et al 1996b\\
6254 (M10) & 69 & $14.63\pm0.04^{a}$ & 0.30 & 71 & 13 & $0.97\pm0.17$ & Harris et al 1976 \\
Rup 106 & 65 & $17.81\pm0.03$ & 0.29 & 46 & 12 & $1.41\pm0.28$ & Buonanno et al 1993 \\
& & & & & & & \\
5694 & 56 & $18.58\pm0.03^{a}$ & 0.27 & 43 & 14 & $1.30\pm0.27$ &
Ortolani \& Gratton 1990 \\
6397 & 105 & $12.88\pm0.04$ & 0.27 & 91 & 10 & $1.15\pm0.17$ &
Kaluzny 1997 \\
6809 (M55) & 203 & $14.43\pm0.02^{a}$ & 0.28 & 197 & 37 &
$1.03\pm0.11^{b}$ & Lee 1977c \\
& & & & & & & \\
4590 (M68) & 108 & $15.64\pm0.01$ & 0.26 & 119 & 34 &
$0.91\pm0.12$ & Walker 1994 \\
5024 (M53) & 302 & $16.89\pm0.01$ & 0.27 & 278 & 39 & $1.09\pm0.09$ & Rey et al. 1998 \\
5053 & 53 & $16.65\pm0.03$ & 0.25 & 58 & 10 & $0.91\pm0.18$ & Sarajedini \& Milone 1995 \\
5466 & 86 & $16.57\pm0.02$ & 0.26 & 71 & 12 & $1.21\pm0.20$ & Buonanno et al 1984a \\
6341 (M92) & 140 & $15.05\pm0.02$ & 0.26 & 92 & 20 & $1.52\pm0.20$ & Buonanno et al 1983b \\
7078 (M15) & 153 & $15.83\pm0.01$ & 0.26 & 124 & 23 & $1.23\pm0.15$ &
Buonanno et al 1983a \\
7099 (M30) & 202 & $14.99\pm0.03$ & 0.27 & 124 & 11 &
$1.63\pm0.19$ & Sandquist et al 1998, \\
& & & & & & & Yanny et al 1994 \\
\end{tabular}
\end{minipage}
\end{table*}

\begin{table*}
\begin{minipage}{15cm}
\contcaption{}
\begin{tabular}{@{}lclcccll@{}}
ID (NGC/IC) & $N_{HB}$ & $V_{HB}$ & $\Delta V_{BC}$ & $N_{RGB}$ & $N_{AGB}$ &
$R$ & References \\
\multicolumn{8}{c}{\underline{Magellanic Clouds Clusters}}\\
1466 & 91 & $19.33\pm0.02$ & 0.28 & 96 & 36$^{c}$ &
$0.95\pm0.14$ & Walker 1992c \\
1841 & 165 & $19.35\pm0.03$ & 0.26 & 158 & 32 & $1.04\pm0.12$ & Brocato et al 1996c \\
2210 & 67 & $19.10\pm0.03^{a}$ & 0.27 & 70 & 15 & $0.96\pm0.17$ & Brocato et al 1996c \\
2257 & 164 & $19.02\pm0.03$ & 0.28 & 138 & 24 & $1.19\pm0.14$ &
Testa et al 1995 \\
Reticulum & 46 & $19.07\pm0.01$ & 0.29 & 54 & 12 & $0.85\pm0.17$ & Walker 1992b \\
\hline
\end{tabular}

\medskip
Notes: $^{a}$~A theoretical correction has been applied to compute the
level of the instability strip.

$^{b}$~The possibility of incompleteness exists at the faint end of
the sample.

$^{c}$~The number may be affected field star contamination.
\end{minipage}
\end{table*}

The variation of the $R$ values as a function of HB type is shown in
Fig.~\ref{rvhb}. There may be a slight decrease in the average $R$ for
clusters with bluer HB type. Several clusters with the bluest
morphologies have unusually high $R$ values: NGC 6752, NGC 7099 (M30),
NGC 6341 (M92), NGC 6218 (M12), and NGC 6681 (M70).  However, this is
not universal --- there are more clusters in the same range of
$R_{HB}$ having $R$ values that are closer to the Galactic
average. There is not a clear reason for this difference:
well-measured clusters are found in both groups, and a metallicity
difference does not appear to be present.

\begin{figure}
\vspace{5.5cm}
\caption{Globular cluster $R$ values versus their horizontal branch
types ($R_{HB}=(B-R)/(B+V+R)$). $\bullet$ indicates clusters having
greater than 200 observed HB stars, $\circ$ indicates clusters having
between 100 and 200, $\ast$ indicates clusters with less than 50, and
$\triangle$ indicates LMC clusters.}
\label{rvhb}
\end{figure}

An important effect of the inclusion of the metallicity dependence of
$\Delta V_{BC}$ is the improved agreement between the $R$ values of several of
the most metal-rich clusters and the metal-poor clusters. In a similar
fashion, the increased $\Delta V_{BC}$ values will also significantly
reduce the estimates of the helium abundances of Galactic bulge fields
($Y=0.28\pm0.02$ according to Minniti 1995) because of the high mean
metallicity of those stars.

\subsection{The MS--HB Magnitude Difference $\Delta$}\label{delta}

The indicator $\Delta$ (CCC) is defined simply as the magnitude
difference between the MS at $(B - V)_{0} = 0.7$ and the HB at the
instability strip. CCC originally defined the HB point to be at the
blue edge of the instability strip, but we have chosen to revise this
definition to make it more easily calculable theoretically and
observationally. This also reduces the color difference between the HB
and MS points, reducing possible systematic effects from photometric
calibration. The various sensitivities of the indicator are not
significantly changed by the revision to the definition.

Increases in envelope helium abundance influence this indicator by
increasing the luminosity of the HB (via the strength of hydrogen
shell-burning during that phase) and by increasing the effective
temperature and the luminosity of the MS through a decrease in the
envelope opacity.  The net effect of the luminosity and temperature
changes on the MS is to make it fainter at a {\it given} color.  As an
indicator, $\Delta$ has the advantage of being sensitive to the helium
abundance ($\partial \Delta / \partial Y = 5.8$~mag; CCC), and the
disadvantages of having a definite metallicity dependence ($\partial
\Delta / \partial \mbox{[Fe/H]} \approx -0.5$ mag / dex) and of requiring
photometry from the HB to well below the MS turnoff.  There is an
additional disadvantage in requiring the knowledge of the cluster's
reddening to determine $V_{0.7}$. However, $\Delta$ has no dependence
on age, since the chosen MS point reaches unevolved stars.  We can
calculate theoretical values of $\Delta$ using a self-consistent set
of isochrones (BV92) and HB models (Dorman 1992) having $Y \approx
0.236$. We have fitted the following polynomial to the data after
making small corrections for different initial helium content in the
theoretical models: \[ \Delta = 4.268 - 2.1295 \mbox{[M/H]} - 0.7938
(\mbox{[M/H]})^{2} \] \[ \hspace{2in} - 0.1173 (\mbox{[M/H]})^3 .\]

As can be seen from Table~\ref{tdelta}, there are only 20 clusters with
enough information to compute a value of $\Delta$.
We included clusters for which $V_{HB}$ and $V_{0.7}$
were derived from different studies, in spite of the possibility of
zero point differences. The error budget for $\Delta$ was computed
using: 
\[ \sigma^{2}(\Delta) = \sigma^{2}(V_{HB}) + \sigma^{2}(V_{0.7}) \]
\[ \hspace{1in} + (\frac{dV_{MS}}{d(B-V)})^{2} \sigma^{2}(E(B-V)) +
\sigma^{2}(\emptyset) \]
where the last two terms account for reddening and zero point
uncertainties respectively. As was discussed in \S~\ref{secr}, we do
not need to worry about how well calibrated the data are in an absolute
sense, as long as there are not significant systematic errors in the
relative photometry for the MS and HB (for instance, a nonlinearity in
one of the CCDs). 

\begin{table*}
\begin{minipage}{15cm}
\caption{Data for the magnitude difference $\Delta$}
\label{tdelta}
\begin{tabular}{@{}llcccll@{}}
ID (NGC) & $V_{HB}$ & $E(B-V)$ & $V_{0.7}$ & $\Delta$ &
HB Reference & MS Reference \\
\hline
6838 (M71) & $14.41\pm0.03^{a}$ & $0.28\pm0.02$ &
$19.80\pm0.12$ & $5.39\pm0.12$ & Hodder et al 1992 & Hodder et al 1992 \\
& & & & & & \\
104 (47 Tuc) & $14.02\pm0.05^{a}$ & $0.04\pm0.01$ &
$19.43\pm0.09$ & $5.41\pm0.07$ & Hesser et al 1987 & Hesser et al 1987 \\
& & & & & & \\
6171 (M107) & $15.62\pm0.02$ & $0.31\pm0.02$ & $20.81\pm0.13$ &
$5.19\pm0.13$ & Ferraro et al 1991 & Ferraro et al 1991 \\
& & & & & & \\
362 & $15.42\pm0.02^{a}$ & $0.04\pm0.01$ & $21.36\pm0.07$ &
$5.94\pm0.09$ & Harris 1982 & Bolte 1987 \\
1261 & $16.72\pm0.03$ & $0.00\pm0.02$ & $22.32\pm0.11$ &
$5.60\pm0.12$ & Alcaino et al 1992 & Alcaino et al 1992 \\
1851 & $16.15\pm0.02$ & $0.02\pm0.02$ & $21.60\pm0.09$ &
$5.45\pm0.09$ & Walker 1992a & Walker 1992a \\
6121 (M4) & $13.50\pm0.09$ & $0.37\pm0.01$ & $18.98\pm0.06$ &
$5.48\pm0.11$ & Kanatas et al 1995 & Richer \& Fahlman 1984 \\
& & & & & & Alcaino et al 1988 \\
& & & & & & \\
288 & $15.26\pm0.04^{a}$ & $0.03\pm0.01$ & $21.17\pm0.07$ &
$5.91\pm0.10$ & Bolte 1992 & Bolte 1992 \\
5139 ($\omega$ Cen) & $14.56\pm0.01$ & $0.15\pm0.02$ &
$20.60\pm0.13$ & $6.04\pm0.16$ & Kaluzny et al 1997 & Alcaino \&
Liller 1987 \\
5904 (M5) & $15.09\pm0.02$ & $0.03\pm0.01$ & $20.88\pm0.06$ &
$5.78\pm0.06$ & Sandquist et al 1996 & Sandquist et al 1996 \\
6752 & $13.70\pm0.05^{a}$ & $0.04\pm0.01$ &
$19.69\pm0.07$ & $5.99\pm0.09$ & Penny \& Dickens 1986 & Penny \&
Dickens 1986 \\
& & & & & & \\
5272 (M3) & $15.66\pm0.03$ & $0.00\pm0.01$ & $21.50\pm0.05$ &
$5.84\pm0.06$ & Ferraro et al 1997 & Ferraro et al 1997 \\ 
6218 (M12) & $14.90\pm0.05$ & $0.22\pm0.04$ & $20.43\pm0.28$ &
$5.53\pm0.30$ & Racine 1971 & Sato et al 1989 \\
6254 (M10) & $14.62\pm0.05^{a}$ & $0.27\pm0.03$ &
$20.68\pm0.12$ & $6.06\pm0.13$ & Hurley et al 1989 & Hurley et al 1989 \\
& & & & & & \\
6397 & $12.97\pm0.05^{a}$ & $0.18\pm0.02$ &
$19.27\pm0.10$ & $6.30\pm0.11$ & Alcaino et al 1987 & Alcaino et al 1987 \\
& & & & & & \\
4590 (M68) & $15.64\pm0.01$ & $0.07\pm0.01$ & $22.00\pm0.05$ &
$6.36\pm0.05$ & Walker 1994 & McClure et al 1987 \\
6341 (M92) & $15.10\pm0.01$ & $0.02\pm0.01$ & $21.40\pm0.05$ &
$6.30\pm0.05$ & Carney et al. 1992b & Stetson \& Harris 1988 \\
7078 (M15) & $15.81\pm0.01$ & $0.10\pm0.01$ & $22.15\pm0.05$ &
$6.34\pm0.05$ & Silbermann \& Smith 1995b & Durrell \& Harris 1993 \\
7099 (M30) & $15.10\pm0.02$ & $0.05\pm0.02$ & $21.61\pm0.10$ &
$6.52\pm0.10$ & Bergbusch 1996 & Bergbusch 1996 \\
\hline
\end{tabular}

\medskip
Note: $^{a}$~$V_{HB}$ has been corrected to the instability strip.
\end{minipage}
\end{table*}

In a global sense, the degree of agreement between the observed values
and theoretical expectations is primarily a function of the
metallicity scale chosen. In Fig.~\ref{dfig}, we show the $\Delta$
values using metallicities from three different studies: ZW, Carretta
\& Gratton (1996; hereafter CG), and Rutledge, Hesser, \& Stetson
(1997; hereafter RSD). The ZW comparison agrees well with the theory,
although the scatter in values at a given metallicity is large. The
high-resolution spectroscopy study of CG reduces the scatter
considerably, although there seems to be a constant offset between the
observed and theoretical values. RSD calibrated their Ca II triplet
measurements to both the ZW and CG scales. In Fig.~\ref{dfig}, we plot
the values using RSD's CG scale. Again, the scatter is fairly low, but
the shape of the curve traced by the observed values is very different
from theoretical expectations. The situation is almost identical for
their ZW-calibrated values.

\begin{figure}
\vspace{5.5cm}
\caption{The $\Delta$ indicator as as function of [Fe/H]. The three
panels use metallicities taken from the studies cited. The Rutledge et
al. panel uses their metallicities calibrated to Carretta \& Gratton's scale.}
\label{dfig}
\end{figure}

How are we to judge the merits of the different comparisons?  First we
must keep in mind that the absolute metallicity scale is uncertain at
the 0.2 dex level (RSD), as has been assumed throughout the paper. The
relative rankings are to be trusted more ($\sigma \sim 0.1$~dex),
making the shape of the $\Delta$ comparison most reliable. So,
horizontal shifts of the curve by 0.2 dex should not be considered
unreasonable. This means that the comparisons using both the ZW and CG
metallicity data are consistent with the theoretical values.  The RSD
scales result in large discrepancies, particularly at the metal-rich
end. The fact that the shape is different in both their ZW and CG
calibrations implies that it probably results from details or
assumptions of their technique. RSD discuss the issue of their Ca
measurements as indicators of [Fe/H] at length, and we refer readers
to their \S~6.

Because the absolute metallicity scale remains uncertain at about the
0.2 dex level, it is currently impossible to determine low-error
absolute helium abundances using $\Delta$. At the metal-rich end in
particular, the absolute metallicity uncertainty causes large
uncertainties in absolute helium abundance. The relative rankings are
to be trusted more if the relative metal abundances for the globular
clusters are good.  This is not as much the case for the ZW scale
(since it was a weighted average of determinations by different
methods) as it is for the CG and RSD studies since they both applied a
single method in a uniform way. We have chosen to use the CG scale
because of the general agreement of the observational and theoretical
curve shapes.  The helium abundance for each of the clusters was
computed relative to the theoretical values given the derivative
$\partial \Delta / \partial Y$ (CCC). In Figs.~\ref{yfe} and
\ref{dyhb}, we plot the calculated $\delta Y$ values as a function of
metallicity and HB morphology, respectively. It is unsettling that the
clusters with $\left| R_{HB} \right| < 0.8$ (and well-defined blue and
red edges to the instability strip) tend to have higher $\delta Y$
values. NGC 288 and M30 stand out slightly as having high values among
the clusters with the bluest morphologies, and NGC 362, M71, and 47
Tuc stand out as high for clusters with red morphologies. This may
indicate that the level of the horizontal branch is not being measured
properly. However, the magnitude error would have to be between about
0.1 and 0.25 mag to bring them back into agreement, which seems
overly large. If the reddest HB clusters can be explained via errors in
their metallicities, there might be slight evidence for increased helium
abundance with bluer HB type. We will discuss discrepant individual
clusters in \S~\ref{comp}.

\begin{figure}
\vspace{5.5cm}
\caption{$\delta Y(\Delta)$ values as a function of $R_{HB}$ (Lee, Demarque, \&
Zinn 1994).}
\label{dyhb}
\end{figure}

\subsection{The RR Lyrae Mass-Luminosity Exponent $A$}

The indicator $A$ (CCC) is related to the mass-luminosity relationship
for stars inside the instability strip: \[ A = \log (L/L_{\odot}) -
0.81 \log (M/M_{\odot}) .\] $A$ is dependent on helium because
increased helium can both increase the luminosity of an RR Lyrae, as
well as the mean mass of stars occupying the instability strip. While
$A$ has a relatively small sensitivity to helium abundance ($\partial
A / \partial Y = 1.4$; Sandage 1990b, Bencivenni et al. 1991),
statistical errors are generally small for clusters with a fair number
of RR Lyraes ($\sigma_{A} \approx 0.01$).  Potentially $A$ could
provide helium abundances with the best precision ($\sigma_{Y} \sim
0.007$) of the three indicators we have considered.

\Aav can be computed using the period relation of van
Albada \& Baker (1971): \[ \log P = 11.497 + 0.84 A - 3.481 \log
T_{eff} .\] One of the difficulties in using this relation is the
computation of realistic effective temperatures for the cluster
variables from readily observable quantities. As will be shown below,
the uncertainty in the absolute temperature values results primarily
in systematic shifts in \Aav values. As a result, \Aav can only be
realistically considered a relative indicator of helium abundance at
this time.

One source of uncertainty in the temperature scale relates to the
calibration samples of field RR Lyraes. Aside from
uncertainty in the model atmospheres used to calibrate the colors, the
optimal choice of color continues to be debated. $(V-K)$ has been
recommended by many authors (Liu \& Janes 1990; Jones et al. 1992;
Carney, Storm, \& Jones 1992a; Fernley 1993) over $(B-V)$ due to evidence of
shock-wave effects on $B$ magnitudes near maximum light. However,
McNamara (1997) showed that the temperature calibration from $(V-K)$
disagreed systematically as a function of period with those from
$(B-V), (b-y), (V-R)$, and $(V-I)$ when 1994 Kurucz
model atmospheres were used.

In recent years, studies of globular cluster RR Lyraes have turned to
the use of quantities like the period $\log P$ and blue amplitude $A_{B}$ to
derive temperatures so as to avoid systematic errors resulting from
reddenings. However, there is still uncertainty in the temperature
zero-point from the model atmospheres used to calibrate the
temperature of field RR Lyraes, and from the differences in
temperatures derived from colors using different filter combinations.

We have chosen to recalibrate the temperature relation of Catelan
(1998) for RR Lyrae stars of type ab (with $A_{B}$ and [Fe/H] the only
variables) to temperatures derived using $(B-V)$ colors. Catelan,
Sweigart, \& Borissova (1998a) point out that the use of a $\log P$
term in the determination of effective temperatures (as in equation 16
of Carney et al. (1992a)) tends to cause luminosity differences among
RR Lyraes to be translated into temperature differences, erroneously
reducing the scatter in the $P - T_{eq}$ plane.

Our decision to use $(B-V)$ color temperatures for the calibrating
sample of field RR Lyraes is based on the findings of McNamara (1997),
and Kov\'{a}cs \& Jurcsik (1997). As mentioned above, McNamara found
that temperatures derived from $(V-K)$ deviated systematically from
several other commonly-used colors as a function of period. It has
been known for a long time that $M_{K}$ correlates strongly with $\log
P$ (e.g.  Longmore et al. 1990).  Because the $K$ band is on the
Rayleigh-Jeans portion of the blackbody curve for RR Lyraes, it is not
very sensitive to temperature ($B$ and $V$ fall near the maximum).
Kov\'{a}cs \& Jurcsik's examination of $M_{V}, M_{I_{c}}$, and $M_{K}$
as a function of Fourier light curve parameters for globular cluster
variables indicates that the period dependence is important for each
of the filters, but is larger for redder filters. As a result, filter
combinations with longer wavelength baselines have larger period
dependences.  In particular, $(V-K)$ is predicted to have a period
dependence that can explain the trend McNamara sees, while $(V-I_{c})$
has a dependence that is over twice as small, and the dependence for
$(B-V)$ is over ten times smaller.  Because period depends
significantly on the stellar luminosity in addition to temperature
(see van Albada \& Baker's pulsation equation), a large period
dependence in the color is likely to be a systematic problem, in
agreement with the assertion of Catelan et al. (1998a).

So, using temperatures from McNamara (1997), pulsational amplitudes
from Blanco (1992), and metallicities from Layden et al. (1996) for
field RR Lyraes, we found the relation: \[\Theta_{eq} =
(0.776\pm0.012) - (0.030\pm0.008) A_{B}  \] \[ \hspace{2in}- (0.008\pm0.003)
\mbox{[Fe/H]} .\] The fit has a multiple correlation coefficient $r =
0.925$, and an rms deviation of 40 K from the fit.  In addition, the
residuals show no correlation with $\log P$. To check the effect that
different compositions for field and cluster RR Lyraes could have, we
redid the fit using only variables with [Fe/H] $< -1.0$.  The fit
(using 15 stars) was \[\Theta_{eq} = (0.786\pm0.020) - (0.039\pm0.011)
A_{B}  \] \[ \hspace{2in}- (0.008\pm0.009) \mbox{[Fe/H]} .\] We have chosen to use this
second calculation for the calculations in this paper, though it makes
only small changes to the \Aav values.

In computing \Aav values for cluster variables, we have chosen not to
include variables which are known to exhibit the Blashko effect, as
this causes changes in pulsation amplitude (and in our computed
temperatures) from cycle to cycle. Those variables were included in
calculations of the average period though, since the period of
pulsation is not affected. Error analysis was carried out on $A$
values for each variable.  However, we find that the error in
individual $A$ values is not very significant compared with the
scatter in $A$ values for variables within a cluster.  From an
examination of histograms of $A$ values for the most populated
clusters, the distributions appear to be Gaussian.

We have found 50 Milky Way globular clusters having good data on at
least 2 RRab variables, for a total of 974 stars. We have also
analyzed 8 old Magellanic Cloud clusters (108 variables), and 4 Local
Group dwarf spheroidal galaxies (214 variables), as shown in
Table~\ref{ta}.  Only 25 of the Milky Way clusters (and 5 in the
Magellanic Clouds) have 10 or more RRab stars with good data.  The
table includes only those clusters for which there is at least one
RRab star with period and amplitude measurements. For the
clusters examined, we give the mean period of the RRab stars in column
(3), along with the error in the mean.  The average mass-luminosity
exponent \Aav is computed from the average of the values for the stars
that have $A_{B}$ values. When this is different from the number that
have periods, the number is given in parentheses in column (2).
Fig.~\ref{caput} plots our data against those calculated according to
the method of Caputo \& De Santis (1992), which employed a different
(but also reddening-independent) method for computing $A$. The primary
difference between the studies is in zero-points, which is probably
due to the differences in the temperature calibrations. There may also
be a slope difference at high values of \Aav.

\begin{figure}
\vspace{5.5cm}
\caption{A comparison of \Aav values from the present study with
$\langle A(12) \rangle$~ values from Caputo \& De Santis (1992).}
\label{caput}
\end{figure}

\begin{table*}
\begin{minipage}{15cm}
\caption{Data for the RRab mass-luminosity relation \Aav}
\label{ta}
\begin{tabular}{@{}lclll@{}}
ID (NGC/IC) & $N_{ab}$ & \Pav & \Aav & References \\
\hline
\multicolumn{5}{c}{\underline{Milky Way Clusters}}\\
104 (47 Tuc) & 1 & 0.737 & $2.039$ & Carney et al 1993 \\
6388 & 4 & $0.746\pm0.071$ & $2.039\pm0.044$ & Hazen \& Hesser 1986,
Silbermann et al 1994\\
& & & & \\
6171 (M107) & 15 & $0.537\pm0.018$ & $1.853\pm0.013$ & Dickens 1970 \\
6366 & 1 & 0.508 & 1.852 & Clement 1997 \\
& & & & \\
6362 & 15 & $0.556\pm0.018$ & $1.874\pm0.012$ & Clement et al 1995a \\
6638 & 1 & 0.666 & 1.931 & Clement 1997 \\
6712 & 7 & $0.557\pm0.022$ & $1.875\pm0.012$ & Sandage et al 1966 \\
6723 & 23(20) & $0.536\pm0.014$ & $1.862\pm0.012$ & Menzies 1974 \\
& & & & \\
362 & 7 & $0.542\pm0.023$ & $1.860\pm0.017$ & Sawyer Hogg 1973 \\
1261 & 13 & $0.555\pm0.013$ & $1.870\pm0.012$ & Wehlau \& Demers 1977 \\
1851 & 21(18) & $0.571\pm0.014$ & $1.873\pm0.014$ & Walker 1998 \\
2808 & 1 & 0.539 & 1.857 & Clement \& Hazen 1989 \\
6121 (M4) & 30(25) & $0.538\pm0.010$ & $1.849\pm0.009$ & Sujarkova \& Shugarov 1981 \\
6266 (M62) & 62(40) & $0.544\pm0.009$ & $1.851\pm0.009$ & Sawyer
Hogg 1973 \\
6402 (M14) & 40 & $0.572\pm0.012$ & $1.886\pm0.009$ & Wehlau \& Froelich 1994 \\
6717 (Pal 9) & 1 & 0.575 & 1.895 & Clement 1997 \\
6864 (M75) & 3 & $0.531\pm0.030$ & $1.842\pm0.022$ & Pinto et al 1982 \\
& & & & \\
288 & 1 & 0.679 & 1.972 & Hollingsworth \& Liller 1977 \\
4499 & 63(59) & $0.580\pm0.008$ & $1.879\pm0.006$ & Walker \& Nemec 1996 \\
5139 ($\omega$ Cen) & 83(74) & $0.646\pm0.011$ & $1.929\pm0.009$ &
Petersen 1994 \\
5904 (M5) & 84(80) & $0.552\pm0.008$ & $1.863\pm0.006$ & Storm et al
1991, Brocato et al 1996a, Caputo et al. 1999 \\
6229 & 11 & $0.527\pm0.013$ & $1.818\pm0.011$ & Sawyer Hogg 1973\\
6235 & 2 & $0.600\pm0.016$ & $1.927\pm0.010$ & Clement 1997 \\
6284 & 6 & $0.588\pm0.023$ & $1.909\pm0.012$ & Clement 1997 \\
6544 & 1 & 0.57 & 1.90 & Hazen 1993 \\ 
6558 & 6 & $0.556\pm0.044$ & $1.899\pm0.032$ & Hazen 1996 \\
6584 & 34 & $0.560\pm0.011$ & $1.876\pm0.007$ & Millis \& Liller
1980 \\
6626 (M28) & 9 & $0.586\pm0.022$ & $1.895\pm0.020$ & Wehlau \& Butterworth 1990 \\
6681 (M70) & 1 & 0.564 & 1.911 & Liller 1983 \\
6715 (M54) & 27 & $0.553\pm0.012$ & $1.856\pm0.011$ & Sawyer Hogg 1973\\
6934 & 45 & $0.561\pm0.012$ & $1.853\pm0.008$ & Sawyer Hogg \&
Wehlau 1980 \\
6981 (M72) & 25 & $0.559\pm0.009$ & $1.851\pm0.008$ & Dickens \&
Flinn 1972 \\
7006 & 53(50) & $0.566\pm0.005$ & $1.861\pm0.005$ & Wehlau et al 1992 \\
7089 (M2) & 13 & $0.623\pm0.020$ & $1.925\pm0.017$ & Sawyer Hogg 1973\\
7492 & 1 & 0.805 & 2.015 & Clement 1997 \\
& & & & \\
1904 & 2 & $0.685\pm0.051$ & $1.952\pm0.027$ & Sawyer Hogg 1973\\
3201 & 72(66) & $0.555\pm0.005$ & $1.860\pm0.004$ & Cacciari 1984 \\
5272 (M3) & 148(92) & $0.553\pm0.006$ & $1.865\pm0.004$ & Sandage 1959 \\ 
5286 & 10 & $0.614\pm0.019$ & $1.915\pm0.016$ & Liller \& Lichten 1978a \\
5897 & 4 & $0.735\pm0.094$ & $1.956\pm0.086$ & Wehlau 1990 \\
5986 & 7 & $0.650\pm0.035$ & $1.945\pm0.019$ & Liller \& Lichten 1978b \\
6093 (M80) & 4 & $0.651\pm0.012$ & $1.931\pm0.013$ & Wehlau et al 1990 \\
6139 & 3 & $0.537\pm0.113$ & $1.893\pm0.059$ & Hazen 1991 \\
6273 (M19) & 1 & 0.507 & 1.838 & Clement \& Sawyer Hogg 1978 \\
6333 (M9) & 8 & $0.622\pm0.018$ & $1.940\pm0.010$ & Clement et al
1984, Clement \& Shelton 1996 \\
6656 (M22) & 9 & $0.640\pm0.015$ & $1.922\pm0.014$ & Wehlau \&
Sawyer Hogg 1978 \\
Palomar 13 & 4 & $0.572\pm0.012$ & $1.879\pm0.006$ & Sawyer Hogg 1973\\
Rup 106 & 12(10) & $0.616\pm0.006$ & $1.883\pm0.002$ & Kaluzny et al 1995 \\
& & & & \\
2298 & 1 & 0.640 & 1.924 & Clement et al 1995b \\
4147 & 4 & $0.531\pm0.029$ & $1.827\pm0.031$ & Newburn 1957 \\
4833 & 7 & $0.685\pm0.031$ & $1.951\pm0.020$ & Demers \& Wehlau 1977 \\
5634 & 3 & $0.621\pm0.019$ & $1.925\pm0.012$ & Liller \& Sawyer Hogg 1976 \\
5824 & 7 & $0.624\pm0.008$ & $1.920\pm0.006$ & Clement 1997\\
6293 & 2 & $0.600\pm0.006$ & $1.933\pm0.008$ & Clement et al 1982 \\
\end{tabular}
\end{minipage}
\end{table*}

\begin{table*}
\contcaption{}
\begin{tabular}{@{}lclll@{}}
ID (NGC/IC) & $N_{ab}$ & \Pav & \Aav & References \\
2419 & 25 & $0.656\pm0.007$ & $1.933\pm0.006$ & Pinto \& Rosino 1977 \\
4590 (M68) & 11(9) & $0.627\pm0.016$ & $1.922\pm0.014$ & Walker 1994 \\
5024 (M53) & 24(23) & $0.646\pm0.014$ & $1.929\pm0.007$ & Goranskij 1976 \\
5053 & 5 & $0.672\pm0.026$ & $1.906\pm0.020$ & Nemec et al 1995 \\
5466 & 11(9) & $0.637\pm0.023$ & $1.921\pm0.018$ & Sawyer Hogg 1973\\
6341 (M92) & 9(6) & $0.628\pm0.017$ & $1.907\pm0.022$ & Carney et al 1992b \\
6426 & 4 & $0.665\pm0.020$ & $1.926\pm0.015$ & Sawyer Hogg 1973\\
7078 (M15) & 31(25) & $0.640\pm0.010$ & $1.918\pm0.009$ & Silbermann \&
Smith 1995, Sandage 1990a\\
7099 (M30) & 3 & $0.698\pm0.026$ & $1.965\pm0.029$ & Sawyer Hogg 1973\\
& & & & \\
\multicolumn{5}{c}{\underline{Magellanic Clouds Clusters}}\\
121 & 4 & $0.548\pm0.031$ & $1.845\pm0.026$ & Walker \& Mack 1988a \\
1466 & 25(21) & $0.581\pm0.013$ & $1.876\pm0.008$ & Walker 1992c \\
1786 & 5 & $0.677\pm0.042$ & $1.942\pm0.022$ & Walker \& Mack 1988b\\
1835 & 21(18) & $0.587\pm0.017$ & $1.899\pm0.015$ & Walker 1993 \\
1841 & 17(15) & $0.676\pm0.013$ & $1.934\pm0.009$ & Walker 1990 \\
2210 & 14(9) & $0.585\pm0.018$ & $1.879\pm0.020$ & Hazen \& Nemec
1992, Reid \& Freedman 1994 \\
2257 & 18(15) & $0.592\pm0.018$ & $1.886\pm0.014$ & Nemec et al
1985, Walker 1989 \\
Reticulum & 22(21) & $0.552\pm0.012$ & $1.855\pm0.007$ & Walker 1992b \\
& & & & \\
\multicolumn{5}{c}{\underline{Dwarf Spheroidals}}\\
Carina & 49 & $0.620\pm0.006$ & $1.930\pm0.005$ & Saha et al 1986 \\
Draco & 115(98) & $0.613\pm0.004$ & $1.888\pm0.003$ & Nemec 1985 \\
Sextans & 26(20) & $0.606\pm0.010$ & $1.894\pm0.009$ & Mateo et al 1995 \\
Ursa Minor & 47 & $0.636\pm0.009$ & $1.922\pm0.007$ & Nemec et al 1988 \\
\hline
\end{tabular}
\end{table*}

Caputo \& De Santis found that their \Aav values showed sensitivity to
the HB type --- clusters with blue HBs ($(B-R)/(B+V+R) > 0.7$) tended
to have significantly higher values for \Aav. Synthetic HB models
predict that only very evolved HB stars are found in the instability
strip for clusters with very blue HB populations. In Fig.~\ref{ahb},
we plot our \Aav values versus horizontal branch type $R_{HB}$. As Caputo \& De
Santis found, there is little scatter among Oosterhoff group I
clusters with a few exceptions (most notably NGC 6229 on the low
end). None of the Oosterhoff group II clusters have \Aav values
consistent with the overwhelming majority of Oo I clusters. (Although
Ruprecht 106 has a \Pav~ closer to those of the Oosterhoff II group,
its lack of RRc variables and its low \Aav value indicate that it
could be an Oo I cluster.)  M14 and M28, the two Oo I clusters with
the bluest HB morphologies, have \Aav values that place them at the
low end of the range populated by Oosterhoff II clusters.

\begin{figure}
\vspace{5.5cm}
\caption{The variation \Aav with horizontal branch type
($R_{HB}=(B-R)/(B+V+R)$). Solid symbols indicate clusters with 7
or more RRab variable stars. (All of the dwarf spheroidals tabulated have
samples this large.)}
\label{ahb}
\end{figure}

There are a number of systems that have \Aav values that are
consistent with being in the Oosterhoff II group and HB types that are
redder than the majority of Oo II clusters. This is partly an
incarnation of the second parameter effect in HB morphology. However,
it seems to rule out the possibility that Oo II RR Lyraes are simply
the result of the instability strip being populated solely by stars
that are evolving toward the AGB. The instability strip is too well
populated in systems like the Ursa Minor dSph and Galactic globular
clusters like M68 and NGC 5466, all of which have populations of red
HB stars in addition to the variables.

For this reason, we have re-examined the dependence of \Aav with
[Fe/H] in Fig.~\ref{apvfe}. Many previous studies (e.g. Sandage 1990b,
Caputo \& De Santis 1992) have noted the apparent linear correlation
within the total globular cluster sample.  If this is truly the case,
then the two Oosterhoff groups should also show the same linear
relation separately.  We derive the following relations: \[ A =
(-0.032\pm0.008) \mbox{[Fe/H]} + (1.821\pm0.012) \] for the 21 Oo I
clusters having 7 or more RRab stars (excluding NGC 6229 as an extreme
outlier) having $-1.69 \leq \mbox{[Fe/H]} \leq -0.99$, and \[ A =
(0.001\pm0.018) \mbox{[Fe/H]} + (1.929\pm0.036) \] for 12 Oo II
clusters with 7 or more RRab stars (excluding $\omega$ Cen due to its
metallicity spread) having $-2.22 < \mbox{[Fe/H]} < -1.58$. If
Ruprecht 106 is removed from the Oo I sample (it appears to be a young
cluster, and it has an unusually low dispersion in the $A$ values of
its variables), we find \[ A = (0.003\pm0.010) \mbox{[Fe/H]} +
(1.868\pm0.015) .\] The two groups individually have slopes that are
significantly shallower than derived from the union of the two samples
($-0.088\pm0.006$), and both are consistent with zero.  Both slopes
are also significantly smaller than predictions from synthetic HB
computations ($-0.027$; Lee, Demarque, \& Zinn 1990). For the Oo I
clusters the small slope is not surprising since the \Aav values are not
predicted to have a dependence on HB type for $R_{HB} < 0.7$. The Oo I
clusters M14 and M28 are the only Galactic globular clusters (with 40
and 9 RRab variables respectively) that can be said to fall in the gap
between the two groups. Both have very blue HB morphologies,
indicating that the relatively high temperatures of the variables are
affecting the \Aav values (Caputo et al. 1993).  More unexpected is
the essentially constant value found for the Oo II clusters which are
expected to be much more sensitive to HB type (Caputo \& De Santis
1992).

\begin{figure}
\vspace{5.5cm}
\caption{The variation of (a) $\log$ \Pav and (b) \Aav with [Fe/H], taken
from the compilation of Djorgovski (1993). Solid symbols
indicate clusters with 7 or more RRab variable stars.}
\label{apvfe}
\end{figure}

An examination of the average periods in Fig.~\ref{apvfe} also indicates
that while there does seem to be a linear relation in the total
sample, if the Oosterhoff groups are considered separately the slope
is significantly smaller.  We have included only those clusters having
7 or more RRab stars. We find
\[ \log P = (-0.019\pm0.010) \mbox{[Fe/H]} + (-0.281\pm0.015) \]
for 20 Oosterhoff I clusters, and
\[ \log P = (-0.028\pm0.015) \mbox{[Fe/H]} + (-0.251\pm0.031) \]
for 13 Oosterhoff II clusters. The slopes are consistent with each
other, and somewhat shallower than predicted by evolutionary models
($-0.05$; Lee, Demarque, \& Zinn 1990).  For the two groups, we find
average periods of 0.556 and 0.643 days, respectively. The offset in
periods between the two Oosterhoff groups almost exactly corresponds
to the difference between the average $A$ values, and so is probably
not due to differences in mean temperature of the variables. From van
Albada \& Baker's relation, this indicates that there is likely to be
a difference in the mean mass and/or luminosity of the RR Lyraes in
the two groups.

While the Oo II clusters with the bluest HBs show large scatter in
\Aav as expected for small samples of RR Lyraes, the clusters with
redder HBs (and generally, with larger numbers of HB stars) -- most
notably M68, but also M15, M53, NGC 5053, NGC 5466, and possibly NGC
2419 -- have the same value to within the errors.  What also makes
these clusters unusual is that if they indeed had high helium
abundances as their higher \Aav values might indicate (whether
primordial or due to the action of a deep mixing mechanism), all of
their HB star distributions would be expected to be bluer. For these
clusters we are left with the possibilities that either high helium is
acting against an even stronger third parameter, or that helium does
not vary and there is a different second parameter.

In short, it would be wise to discard most of the Oo II clusters since
they are prone to small number statistics, and so are overly
sensitive to the HB type, making them useless as helium abundance
tracers. However, the variable-rich Oo II clusters stand out in the
constancy of their \Aav values as a group. Based on these data alone a
high helium abundance could be a possible explanation. However, the
data from other indicators show no evidence to support this. It is
clear that some other factor is necessary to completely explain the
large numbers (a total of 104 known for the 6 clusters) of RR Lyraes
in these clusters, but the question is beyond the scope of this paper.
One result of this discussion is that there appears to be no
relationship between the Oosterhoff dichotomy and the second parameter
problem, given the lack of variation in \Aav with HB type.

In the range $-1.8 < \mbox{[Fe/H]} < -1.55$, there are clusters of
both Oosterhoff types with at least moderate numbers of
variables. Three of the Oo I clusters (M3, NGC 3201, and NGC 7006) are
among the most populated with RR Lyraes, and have moderately blue HB types.
The Oo II clusters (M2, M9, M22, and NGC 5286) are among the bluest 
HB types in the group. The four Magellanic Cloud clusters that have \Aav 
between the two Oosterhoff groups also have [Fe/H] $\approx -1.8$
and have HB types falling between the majority of the clusters in the
the two groups.

The difference of 0.06 in $A$ between Oosterhoff I and II clusters
would require a difference of 0.05 in initial helium abundance. A
sudden change of that magnitude is unlikely to have occurred in
Galactic chemical evolution. Enrichment of the envelopes of the RR
Lyraes by deep mixing on the upper RGB could potentially be invoked to
explain the $A$ measurements.  In order to explain observed variations
in the abundances of species of aluminum and magnesium in several
clusters, surface material must have mixed through regions where
hydrogen has at least partly burned to helium in the shell source.
Sweigart (1997) finds that an increase in envelope helium abundance by
0.04 could explain the period shift difference between M3 and M15.
(Period shifts go like 0.84 times the difference in $A$.) If so, the
Oosterhoff dichotomy would imply that the driving mechanism only
existed in one group or the other (in the case of stellar rotation
driving circulation, Oo II clusters would be required to have higher
average rotation). However, significant variations in oxygen abundance
are seen in giants in clusters of both Oo I (M3, Kraft et al. 1992;
M5, Sneden et al. 1992) and Oo II (M15, Sneden et al. 1997; M92,
Sneden et al. 1991) groups, making this explanation unlikely. Among
the Oo I clusters, the $A$ indicator shows no evidence of significant
helium abundance variations.

We have also examined old Magellanic Cloud globular clusters and Local
Group dwarf spheroidals (dSph) having RR Lyrae observations in the
literature. Several of the dSph galaxies have populations that are old
enough to make a comparison with globular clusters useful, though
composition variations have been seen in Draco (Lehnert et al.  1992;
Shetrone, Bolte, \& Stetson 1998) and Sextans (Suntzeff et al.  1993).
In spite of the composition variations, Draco's variables have $A$
values that are strongly peaked near its \Aav, giving additional
evidence for lack of a significant metallicity dependence in A.  

Ursa Minor, the dSph with the largest mean period of those we have
data for, falls unequivocally among Oo II clusters.  Carina, Draco,
and Sextans have mean periods that put them just on the Oo II side of
the period gap between the Oosterhoff groups. However, only Carina and
Ursa Minor have \Aav values consistent with Oo II clusters.  (Our
estimated $R_{HB}$ value for Carina is uncertain to probably $\pm
0.2$, but this does not impact the analysis. The value was estimated
from the appearance of the old population HB in the CMD of Smecker
Hane et al.  1994.) Draco and Sextans have \Aav values slightly higher
than the average for Oo I clusters.  Draco is unusual due to being
very metal poor with an unusually red morphology.  To check if this
affected the derived \Aav values, we derived the following relation
from the sample of 39 globular clusters (Milky Way and LMC) having
more than 7 RR Lyrae stars: \[ \langle \log T_{eff} \rangle = (3.8394
\pm 0.0012) + (0.0100 \pm 0.0008) \mbox{[Fe/H]} \] \[ \hspace{2in}+
(0.0032 \pm 0.0005) R_{HB}.\] The rms residual for the globular
clusters was 0.0025. The dwarf spheroidals do not show good evidence
for deviating from this rough relation, as shown in
Fig.~\ref{tres}. Thus, the metallicity and HB morphology seem to be
able to account for the low temperatures of the RR Lyraes, with no
obvious distiinction between Oosterhoff groups.  As was found for the
reddest Galactic Oo II clusters, helium cannot be the main factor
determining the HB morphology since increased envelope helium would
drive the morphology much bluer.

\begin{figure}
\vspace{5.5cm}
\caption{Residuals for $\langle log T_{RR} \rangle$ (in the sense of
calculated minus observed) as a function of [Fe/H] and $R_{HB}$ for
the globular clusters and dwarf spheroidals having more than 7 RR
Lyraes with good data.}
\label{tres}
\end{figure}

The slope derived from the 5 RR Lyrae-rich Magellanic cloud clusters
is \[ \partial A / \partial \mbox{[Fe/H]} = -0.141\pm0.019 .\] This is
significantly higher than the slope of the (Oo I + Oo II) Galactic
samples, but these clusters might also be more profitably put into
separate Oosterhoff classes.  According to the mean periods, the SMC
cluster NGC 121 and the LMC cluster Reticulum are Oo I clusters, and
NGC 1786 and 1841 are Oo II.  (We have estimated $R_{HB}=0.9\pm0.1$
for NGC 1786 from the statistically subtracted CMD of Brocato et
al. 1996c.) The remaining four LMC clusters (NGC 1466, 1835, 2210, and
2257) fall between the two groups in Figs.~\ref{apvfe}, while having
mean periods at the high end of the Oo I group. It may be suggestive
that all four of these clusters have [Fe/H]$\approx -1.8$, which
appears to be at the metal-poor end of the metallicity distribution
for Oo I clusters in the Milky Way. A high helium abundance could be
the cause only if helium enrichment occurs by a process that did not
enrich the heavy elements in these clusters.

\section{Comparisons}\label{comp}

As stated previously, we have examined these three helium indicators
in an attempt to see if we could determine with greater confidence
whether any of the Galactic globular clusters have abnormal
helium abundances. It is also clear that we are unlikely to derive
useful {\it absolute} helium abundances. We will now see if we can
find anomalous relative abundances under the assumption that abnormal
indicator values reflect abnormal helium abundances (which may not be
the case, given the factors that can affect each indicator).

For $R$, helium abundances can be computed from equation 11 of Buzzoni
et al. (1983). For $\Delta$, we compute
a relative helium abundance value $\delta Y(\Delta)$ using the
theoretical values in \S~\ref{delta} and the derivative $\partial
\Delta / \partial Y = 5.8$~mag (CCC). Here, the uncertainties in the
absolute metallicity scale prevent the computation of good absolute
helium abundances. We have chosen to use CG abundances from
high-dispersion spectroscopy, supplementing with abundances from RSD
for NGC 1261, 1851, and 6218 (M12). For \Aav, we compute relative
helium abundances $\delta Y(A)$ by computing the difference from the
average \Aav value for the corresponding Oosterhoff group, and then
applying the derivative $\partial A / \partial Y = 1.4$ (Sandage
1990b; Bencivenni et al. 1991). Because the
\Aav values do not show significant dependences on metallicity, we
believe this is the best choice. We have chosen not to compute $\delta
Y$ for clusters having fewer than 7 RRab stars with $B$ amplitudes.
For the Oo I clusters, the average value is $1.8629\pm0.0017$, and
$1.9271\pm0.0028$ for Oo II clusters.  The helium values are given in
Table~\ref{ty}, and the comparison plots are shown in Fig.~\ref{yrvs}.

\begin{table}
\caption{Helium abundance comparisons}
\label{ty}
\begin{tabular}{@{}lccc@{}}
ID (NGC/IC) & $Y(R)$ & $\delta Y(\Delta)$ & $\delta Y(A)$ \\
\hline
\multicolumn{4}{c}{\underline{Milky Way Clusters}}\\
5927 & $0.139^{+0.020}_{-0.023}$ & & \\
6496 & $0.220^{+0.028}_{-0.034}$ & & \\ 
6624 & $0.120^{+0.019}_{-0.022}$ & & \\
6637 (M69) & $0.214^{+0.021}_{-0.025}$ & & \\
& & & \\
6838 (M71) & & $0.044\pm0.021$ & \\
104  (47 Tuc) & $0.216^{+0.013}_{-0.015}$ & $0.048\pm0.012$ & \\
6171 (M107) & $0.151^{+0.020}_{-0.023}$ & $-0.046\pm0.023$ &
$-0.007\pm0.009$ \\
6362 & & & $0.008\pm0.009$ \\
6652 & $0.210^{+0.027}_{-0.032}$ & & \\
6712 & $0.206^{+0.021}_{-0.024}$ & & $0.009\pm0.009$ \\
6723 & $0.217^{+0.023}_{-0.027}$ & & $-0.001\pm0.009$ \\
& & & \\
362 & $0.197^{+0.023}_{-0.027}$ & $0.047\pm0.016$ & $-0.002\pm0.012$ \\
6266 (M62) & $0.245^{+0.020}_{-0.023}$ & & $-0.009\pm0.006$ \\
1261 & $0.218^{+0.020}_{-0.022}$ & $0.001\pm0.021$ & $0.005\pm0.009$ \\
6121 (M4) & $0.179^{+0.021}_{-0.024}$ & $-0.039\pm0.019$ & $-0.010\pm0.006$ \\
1851 & $0.206^{+0.016}_{-0.018}$ & $-0.016\pm0.016$ & $0.007\pm0.010$ \\
2808 & $0.175^{+0.015}_{-0.016}$ & & \\
6402 (M14) & & & $0.017\pm0.006^{a}$ \\
288  & $0.203^{+0.023}_{-0.027}$ & $0.056\pm0.017$ & \\
5904 (M5) & $0.188^{+0.010}_{-0.011}$ & $0.027\pm0.010$ & $-0.002\pm0.005$ \\
6584 & & & $0.009\pm0.005$ \\
7006 & $0.248^{+0.025}_{-0.030}$ & & $-0.001\pm0.004$ \\
7089 (M2) & & & $-0.002\pm0.012$ \\
& & & \\
4499 & $0.219^{+0.020}_{-0.023}$ & & $0.012\pm0.004$ \\
6229 & $0.162^{+0.022}_{-0.026}$ & & $-0.032\pm0.008$ \\
6626 (M28) & & & $0.023\pm0.014^{a}$ \\
6681 (M70) & $0.258^{+0.027}_{-0.033}$ & & \\
6715 (M54) & & & $-0.005\pm0.008$ \\
6752 & $0.250^{+0.018}_{-0.020}$ & $0.017\pm0.016$ & \\
6934 & $0.156^{+0.028}_{-0.034}$ & & $-0.007\pm0.006$ \\
6981 (M72) & $0.140^{+0.030}_{-0.037}$ & & $-0.009\pm0.006$ \\
6254 (M10) & $0.171^{+0.026}_{-0.031}$ & $0.031\pm0.023$ &  \\
& & & \\
1904 (M79) & $0.187^{+0.021}_{-0.024}$ & & \\
3201 & $0.205^{+0.018}_{-0.020}$ & & $-0.002\pm0.003$ \\
6218 (M12) & $0.256^{+0.026}_{-0.031}$ & $-0.022\pm0.052$ & \\
5272 (M3) & $0.204^{+0.011}_{-0.012}$ & $0.002\pm0.010$ & $0.002\pm0.003$ \\
5286 & & & $-0.009\pm0.011$ \\
5897 & $0.227^{+0.029}_{-0.035}$ & & \\
5986 & & & $0.013\pm0.014$ \\
6093 (M80) & $0.172^{+0.018}_{-0.020}$ & & \\
6205 (M13) & $0.180^{+0.023}_{-0.027}$ & & \\
6333 (M9) & & & $0.009\pm0.007$ \\
6656 (M22) & & & $-0.004\pm0.010$ \\
Rup 106	& $0.233^{+0.029}_{-0.036}$ & & $0.014\pm0.001$ \\
\end{tabular}
\end{table}

\begin{table}
\contcaption{}
\begin{tabular}{@{}lccc}
ID (NGC/IC) & $Y(R)$ & $\delta Y(\Delta)$ & $\delta Y(A)$ \\
4833 & & & $0.017\pm0.014$ \\
5694 & $0.220^{+0.031}_{-0.038}$ & & \\
5824 & & & $-0.005\pm0.004$ \\
6397 & $0.200^{+0.023}_{-0.026}$ & $0.030\pm0.019$ & \\
6809 (M55) & $0.181^{+0.016}_{-0.018}$ & & \\
& & & \\
2419 & & & $0.004\pm0.004$ \\
5024 (M53) & $0.190^{+0.013}_{-0.014}$ & & $0.001\pm0.005$ \\
4590 (M68) & $0.160^{+0.021}_{-0.024}$ & $0.027\pm0.009$ & $-0.004\pm0.010$ \\
7099 (M30) & $0.257^{+0.018}_{-0.021}$ & $0.061\pm0.017$ &  \\
7078 (M15) & $0.211^{+0.019}_{-0.021}$ & $0.014\pm0.009$ & $-0.007\pm0.006$ \\
5466 & $0.208^{+0.025}_{-0.029}$ & & $-0.004\pm0.013$ \\
6341 (M92) & $0.245^{+0.021}_{-0.024}$ & $0.005\pm0.009$ & \\
5053 & $0.161^{+0.029}_{-0.035}$ & &  \\
\multicolumn{4}{c}{\underline{Magellanic Clouds Clusters}}\\
Reticulum & $0.150^{+0.030}_{-0.037}$ & & $-0.006\pm0.005$ \\
1835 & & & $0.026\pm0.011$ \\
2210 & $0.169^{+0.026}_{-0.031}$ & & $0.012\pm0.014$ \\
2257 & $0.204^{+0.018}_{-0.021}$ & & $0.017\pm0.010$ \\
1466 & $0.167^{+0.023}_{-0.026}$ & & $0.009\pm0.006$ \\
1841 & $0.183^{+0.018}_{-0.020}$ & & $0.005\pm0.006$ \\
\multicolumn{4}{c}{\underline{Dwarf Spheroidal Galaxies}}\\
Carina & & & $0.002\pm0.004$ \\
Draco & & & $0.018\pm0.002$ \\
Sextans & & & $0.022\pm0.006$ \\
Ursa Minor & & & $-0.004\pm0.005$ \\
\hline
\end{tabular}

\medskip
Note: $^{a}$~This cluster's \Aav value has probably been affected by
its HB morphology, and so the $\delta Y$ value should be regarded as
an upper limit.
\end{table}

\begin{figure}
\vspace{5.5cm}
\caption{Correlations between helium abundances derived from indicators
$R$ and a) $\Delta$, or b) \Aav. $\triangle$ indicates LMC clusters.}
\label{yrvs}
\end{figure}

There isn't an obvious correlation in the comparison of helium
abundances from the $R$ and $\Delta$ or \Aav indicators.
Comparison between abundances from $\Delta$ and \Aav is hampered by
the small overlap between the two samples. We can alternately look for clusters
whose values may indicate anomalous helium abundances.

In the following paragraphs, we briefly discuss clusters with unusual
values for some or all of the helium indicators.

{\it NGC 1851:} The value
derived from $\Delta$ is below average, and the value from $R$ is about
average. From previous studies of its RR Lyraes (Wehlau et al. 1978,
Wehlau et al. 1982), NGC 1851 appeared to have an \Aav closer to those of
Oosterhoff II clusters. In fact, both the mean period of the variables and the
ratio of RR Lyrae numbers $N_{c} / N_{ab}$ are high for an Oosterhoff
I cluster, and the bimodality of the HB (with a sparsely populated
instability strip) has been difficult to model using canonical HB models
(see Catelan et al. 1998b). However, Walker's (1998) re-examination of the
RR Lyraes using CCD photometry indicates that the $B$-amplitudes in
the photographic studies were systematically too large. The new data
indicates that NGC 1851's \Aav value is completely consistent with the
Oo I average. 

{\it NGC 5927:} see NGC 6624.

{\it M4 (NGC 6121):} All three indicator values for M4 are below the
averages, although the \Aav value is only slightly so. The value
derived from $\Delta$ can be questioned due to indications that there
is differential reddening across the cluster (Cudworth \& Rees 1990),
and that $R_{V} = A_{V} / \mbox{E}(B-V)$ differs from the most
frequently used value (e.g., Dixon \& Longmore 1993). Because the
magnitudes of the HB and MS points were taken from studies of
different portions of the cluster, we might expect a systematic error
of over 0.01 in $Y$.  The reddening quoted in Table~\ref{tdelta} is a
weighted mean of previous studies (Dixon \& Longmore 1993),
although the error of $\pm0.01$ is probably still an underestimate.
The values from $R$ and \Aav should be more reliable.

{\it NGC 6171 (M107):} M107 has very low $Y(R)$ and $\delta Y(\Delta)$
values, and an average $\delta Y(A)$ value. To completely explain the
low $\Delta$ value by a reddening error, E$(B-V)$ would have to be
underestimated by 0.06, which is possible given that E$(B-V) \sim
0.3$. Similarly, to explain the low $R$ value, one would have to
invoke an excessively large metallicity error.

{\it NGC 6229:} Like NGC 1851, NGC 6229 also has a bimodal HB with a
sparsely populated instability strip. The \Aav value derived for this
cluster is considerably {\it lower} than the average. As was
previously the case with NGC 1851 (see above), the RR Lyrae data is
based on fairly old photometry, which may be the source of systematic
error in the pulsational amplitudes. The $R$ value also indicates a
low helium abundance for a fair-sized sample ($\approx 100$ each of
HB and RGB stars). NGC 6229 has a rather extended blue HB morphology 
for an outer halo clusters, contrary to what would be expected from
the \Aav and $R$ values (Borissova et al. 1997).

{\it M62 (NGC 6266):} Like other post-core-collapse clusters (see NGC
6752 and M30), M62 has a rather large $R$ value, although its \Aav value
is fairly close to the average. M62 has heavy differential reddening,
which may affect the determination of the HB magnitude.

{\it M92 (NGC 6341):} M92's $R$ value seems to be unusually large,
although examination of more recent wide-field CCD photometry
indicates that our value is probably too high (Bolte \& Roman 1999; in
preparation). The $\Delta$ value falls a little lower than the
average.

{\it NGC 6624:} This cluster has the lowest $R$ value of any cluster
examined here. Because $\Delta V_{BC}$ is sensitive to [M/H] at the
metal-rich end, and because NGC 6624 is one of the most metal rich
clusters in our sample, the metal abundance is a natural suspect.  NGC
6624 appears to be approximately $0.2-0.4$ dex more metal-rich than 47
Tuc according to Ca II triplet measurements (RSD; Idiart, Thevenin, \&
de Freitas Pacheco 1997; Armandroff \& Zinn 1988), but less than 0.1
dex more metal-rich according to spectrophotometry (Gregg 1994). If we
optimistically take NGC 6624's metallicity to equal that of 47 Tuc, the
$R$ value increases to $0.83\pm0.10$, which does not alleviate the
problem. It was suggested by Richtler et al.  (1994) that the total
metal abundance [M/H] for the cluster is considerably lower than its
iron abundance ([Fe/H]$=-0.37$; ZW) would indicate, based on
comparisons between photometry and BV92 models. This could be the case
if Type I supernovae became the primary source of heavy element
enrichment. If the iron abundance did not actually trace the total
metallicity for this cluster (and it is in fact more metal-poor), this
would mean the the differential bolometric correction we used was too
large, which would help explain the low $R$ value. The cluster NGC
5927 also shows an extremely low $R$ value consistent with this idea
(Ca II triplet measurements indicate that NGC 5927 is about 0.4 dex more
metal-rich than 47 Tuc).  However, M69 and NGC 6496, which are only
$0.1-0.2$ dex more metal poor according to [Fe/H] measures, have $R$
values which are closer to the average for our sample.

{\it NGC 6752:} This cluster presents a rather large value for $R$
from a large sample of stars. The $\Delta$ value is more
consistent with the average of the globular cluster system, although
the HB magnitude is notoriously difficult to determine due to
a complete lack of stars on the instability strip and red HB.

{\it M30 (NGC 7099):} In its high $R$ and $\Delta$ values, M30
presents evidence for a helium abundance enhancement. The
cluster's reddening is under some dispute though. Reducing the
reddening to E$(B-V)=0.02$ would lower the enhancement computed from
$\Delta$ to 0.033, which is consistent with the bulk of the other
clusters.  The $R$ measurement is more certain to be unusually high
given the relatively large bright star samples.

{\it Ruprecht 106:} One concern in photometric analyses of Ruprecht
106 has been the uncertainty in its metallicity.  Francois et al.
(1997) have found [Fe/H]$=-1.6\pm0.25$ from high-resolution
spectroscopy. The metallicity uncertainty plays a small role here
though because of the small dependences of $\Delta V_{BC}$ (at this
low metallicity) and $A$ on [Fe/H]. The sample of stars for $R$ is
relatively small, so $Y(R)$ is consistent with the values for other
clusters to within the errors. Its RR Lyraes have low $A$ values for
the metallicity of the cluster though, which may be a result of the
apparent youth of the cluster (Buonanno et al. 1993), and relatively
high masses of the variables.

\section{Discussion}

In Fig.~\ref{yfe}, we plot the helium abundances derived from the
three indicators as a function of metallicity.  With the current
dataset, we find that the three helium indicators $R$, $\Delta$, and
$A$ now yield trends of helium abundance as a function of metallicity
that are consistent with zero to within the errors over the range of
[Fe/H] sampled.

\begin{figure}
\vspace{5.5cm}
\caption{Helium abundances versus [Fe/H] (Djorgovski 1993) for the
three indicators. a) $R$: $\bullet$ indicates clusters having
greater than 200 observed HB stars, $\circ$ indicates clusters having
between 100 and 200, $\ast$ indicates clusters with less than 50, and
$\triangle$ indicates LMC clusters. b) $\Delta$. c) \Aav.
Points have the same meaning as in Figs.~\ref{ahb} and \ref{apvfe}.}
\label{yfe}
\end{figure}

There does not seem to be an obvious trend in the $Y(R)$ values as a
function of metallicity, although this could be masked by the
considerable scatter in the points. The measurement error is 
primarily the result of Poisson errors. The situation will probably
improve with the careful examination of the HB and RGB populations of
clusters with the largest evolved star populations. A total RGB and HB
star sample in excess of 1000 stars is necessary to reduce the error
bars for individual clusters to $\sigma (Y) \approx 0.01$.

By separating the clusters into three subsamples according to the
total numbers of HB and RGB stars they have, we derive the mean values
listed in Table~\ref{rav}. The standard
deviations in each case are larger than the average measurement error
for the clusters in each subsample, indicating that the scatter in the
measurements may be real. To test this idea, we did Monte Carlo
simulations using the measurement errors to set $Y(R)$ for each
cluster (with the average value for each cluster always centered on
the mean of the sample), and determined the probability of getting a
standard deviation as large as what is observed. For the 17 clusters
with $N_{RGB}+N_{HB}>250$, we find a standard deviation of 0.033 in
$Y$, and an expected standard deviation of 0.018, which results in a
probability less than $10^{-4}$ that the measurement errors can
explain the standard deviation. For the sample with
$N_{RGB}+N_{HB}<200$ (having larger Poisson errors), the corresponding
probability is 0.06.

\begin{table}
\label{rav}
\caption{Helium abundance derived from $R$ as a function of
$N=N_{RGB}+N_{HB}$.}
\begin{tabular}{@{}lccc}
Sample Size & $Y$ & Std. Dev. & $N_{clusters}$ \\
\hline
$N \geq 300$ & $0.199 \pm 0.010$ & 0.036 & 12 (2 LMC) \\
$200 \leq N \leq 300$ & $0.193 \pm 0.009$ & 0.031 & 12 \\
$100 \leq N \leq 200$ & $0.201 \pm 0.007$ & 0.036 & 23 (3 LMC) \\
\hline
\end{tabular}
\end{table}

The derived mean abundance is considerably lower than the expected
primordial value $Y_{P}=0.23$. Because we believe we have removed the
most important (known) systematic errors in the $R$ measurements, we
should ask what might cause the $Y(R)$ values to be low. Brocato,
Castellani, \& Villante (1998) suggest that the uncertainties in the
rate of the $^{12}$C($\alpha,\gamma$)$^{16}$O reaction lead to an
uncertainty in the derived helium abundance of about 0.02. Another
possibility is that there is a systematic effect throwing the
$\Delta V_{BC}$ values off.

In lieu of systematic errors in $\Delta V_{BC}$, we can ask what
physical processes can slow RGB evolution. Langer, Bolte, \&
Sandquist (1999) have suggested that deep mixing processes on the
upper RGB may result in lengthened or reduced evolutionary times for
those stars, which could modulate the resulting $R$ values. This kind
of scenario would mean that the helium abundances measured by this
method (as well as those using HB stars in some way) would be affected
since some of the helium produced by the hydrogen-burning shell would
be mixed into the envelope of the stars that undergo the process.

We can also attempt to measure the trend in $\delta Y(\Delta)$ as a
function of [Fe/H]. The primary difficulty is the clusters at
[Fe/H]$\sim -1.3$ (essentially on the ZW scale) with abnormally low
values. There are good reasons for removing M4 and M12 from
consideration due to differential reddening and large reddening
uncertainty. Along with M4 (which has a relatively low $\Delta$ value
also), M12, M107, and NGC 1851 were the clusters for which we had to
use metallicities from the RHS study since they were not observed by
CG.  We find from clusters with $\Delta$ measurements (excepting M4,
M12, M107, and NGC 1851) has a linear trend: \[ \delta Y(\Delta) =
(0.016\pm0.007) \mbox{[Fe/H]} + (0.049\pm0.011) .\] The slope has
marginal significance, but systematic error in the zeropoint of the
metallicity scale affects the slope of the best-fit line since the
$\delta Y(\Delta)$ values of high metallicity clusters are modified
to a larger degree by metallicity shifts than low metallicity
clusters. If we reduce the metallicity of all clusters by 0.21 dex
(equivalent to ignoring $\alpha$ elements), we derive a best-fit
slope of $-0.010\pm0.007$ --- a change in sign. Thus, the dependence
of helium abundance on [Fe/H] (in addition to the absolute helium
abundance) as derived from $\Delta$ will not be certain until the
absolute metallicity scale is improved.

The data from the $A$ indicator for the two Oosterhoff groups
are individually consistent with constant helium abundance as a function
of [Fe/H]. It is unlikely that the difference in \Aav between the two
groups is due to a difference in helium abundance since a similar jump
does not appear in the other indicators.

To explain the apparent difference in $A$ values between Oosterhoff
groups, there must either be some combination of a mean luminosity or
a mean mass difference between the RR Lyrae stars in the two groups.
If the difference were entirely due to a difference in mean mass,
there would be no observable effect in either the $\Delta$ or $R$
indicators. If there was only a difference in the mean luminosity (the Oo
II clusters RR Lyraes would have to be about 15\% more luminous than
Oo I variables), Oo II clusters should have $R$ values that are 13\%
smaller (since $R$ requires the HB magnitude to compute the faint
limit of the RGB sample), and $\Delta$ values that are higher by about
0.15 mag. From 15 Oo I clusters having more than 5 variables as well
as $R$ values, we find $\overline{R} = 1.100 \pm0.033$. For the 7 Oo II
clusters meeting the same criteria, we find $\overline{R} = 1.139 \pm
0.060$. The values for the two groups are consistent to within the
errors. For $\Delta$, we find $\overline{\delta Y(\Delta)} = 0.0063 \pm
0.0054$ and $0.0153 \pm 0.005$ for 7 Oo I and 3 Oo II clusters
respectively. So again, we have no evidence for helium abundance
differences between the groups, but the small numbers of clusters with
$\Delta$ values makes this a weak comparison.

The difference in mean $A$ value between the groups corresponds to a
difference in the mean RR Lyrae mass of approximately 20\%. This is
not so large as to make it unreasonable that there might be
differences in the amount of mass loss at the tip of the RGB between
the two groups. The observations require the Oosterhoff II clusters to
have variables of lower mass. This goes in a direction opposite what
is needed to explain the HB morphologies of at least some of the
clusters.

Clearly this does not speak to the exact cause of the Oosterhoff
dichotomy, but it can give a little guidance on the details of how the
dichotomy is brought about.

\section{Conclusions}

We begin this section by summarizing what we consider to be the most
important results of the surveys tabulated here. 

With the corrected differential bolometric corrections (larger for
more metal-rich populations), we now see no significant evidence for
variation in the indicator $R$ as a function of metallicity or
horizontal branch type. In particular, metal-rich ([Fe/H] $\leq -1$)
Galactic globular clusters now are more consistent with the mean, and
Galactic bulge fields (Minniti 1995) are also likely to show lower
helium abundances. However, only a handful of clusters have helium
values $Y(R)$ consistent with the favored primordial value $Y_{P} \sim
0.23$.  There is evidence that there is real scatter in the $R$
values, over and above what can be chalked up to measurement errors.
Clusters with anomalously high (M30, M62, and NGC 6752) and low (M68,
M107, NGC 5927, NGC 6229, and NGC 6624) values appear over a range of
metallicity.

For $A$, we find that virtually all of the Oosterhoff I clusters (and
the LMC cluster Reticulum) have values consistent with a constant
helium abundance. The Oosterhoff II clusters with the largest number
of RR Lyrae variables also appear to have a constant \Aav value (as do
the Ursa Minor and Carina dwarf spheroidals, and the LMC globular
cluster NGC 1841). The remaining LMC globular clusters, and the dwarf
spheroidals Draco and Sextans have \Aav values slightly above the
average for the Oo I group. For the richest Oo II systems, evolution
is unable to explain the numbers of RR Lyrae variables, and a
systematic offset in helium abundance also seems unlikely given the
evidence from the other two indicators. The lack of any correlation
between HB type and \Aav is indication that the second parameter
problem is completely independent from the Oosterhoff dichotomy.
The $A$ measurements give us the best constraint on the chemical
evolution parameter $\Delta Y / \Delta Z$. Using the $1 \sigma$ error
bar on the slope of the $A$ -- [Fe/H] relation, we get a limit $\left| \Delta
Y / \Delta Z \right| \la 10$, which is consistent with measurements from
extragalactic HII regions (Pagel et al. 1992).

For $\Delta$, our results are consistent with constant helium abundance
across the range of metallicity sampled. The absolute value of the
helium abundance as well as the exact value of $dY / d(\mbox{[Fe/H]})$
depend on the metallicity scale used. The most discrepant
clusters can potentially be explained by errors in the cluster reddenings.

Comparing data from different indicators, we find that the mean trends
are all consistent with constant helium abundance for metallicities
[Fe/H] $\la -0.7$. We have examined the three indicators so as to use
the information to either bolster or dispute claims of unusual helium
abundance from just one indicator. In examining the clusters with
helium abundances from more than one indicator, we have not found
convincing evidence that any have abnormal helium
abundances. Systematic effects clearly appear to varying degrees in
the data for all three indicators, but we have not been able to
determine the cause in all cases.

We must note that none of the three indicators we have tabulated has
much data covering the most metal-rich clusters where evidence of
helium enrichment may still reside. In general, the photometry for
these clusters is most subject to field star contamination and heavy
reddening, making interpretation difficult. For $R$, there is the
additional problem that the red HB begins to overlap the RGB in the
CMD, making clean values impossible. The redness of the HB also
tends to preclude the possibility of RR Lyrae stars, and hence the
possibility of computing $A$. Reddening and difficulties in finding
the true HB level makes $\Delta$ values unlikely without considerable
work. Other means must be devised to determine good helium abundances
for these clusters.

\section{Acknowledgements}

I would like to thank many for supplying electronic copies of their
datasets: E. Brocato, F. Ferraro, P. Guhathakurta, D. Martins,
S.-C. Rey, T.  Richtler, N. Samus, A. Sarajedini, V. Testa, and
A. Walker. I would also like to thank M. Catelan, G. E. Langer,
R. Taam, and the anonymous referee for very helpful conversations, and
M. Bolte and R. Taam for their support (under NSF grants AST-9420204
and AST-9415423, respectively) while this work was in progress.

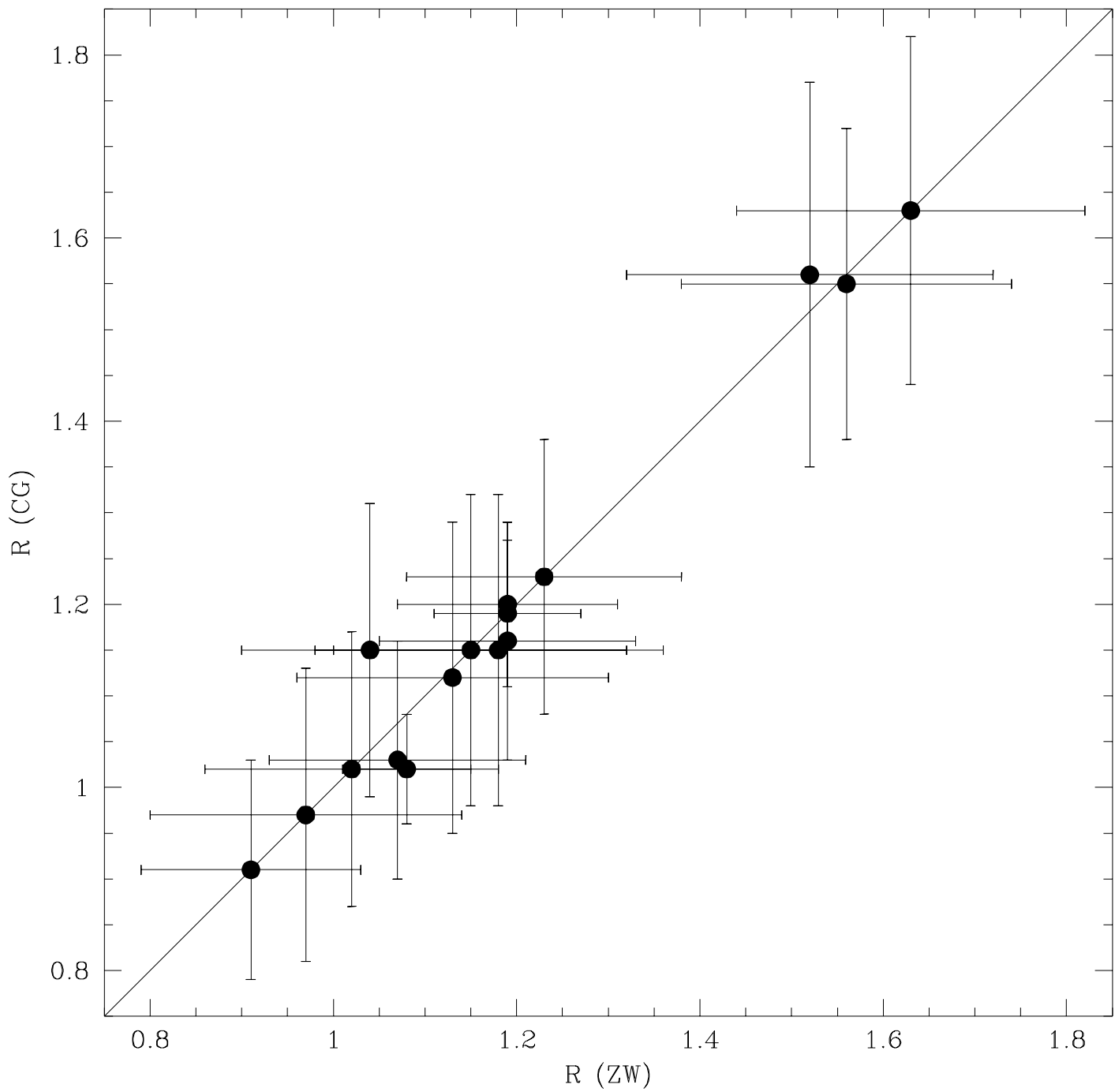
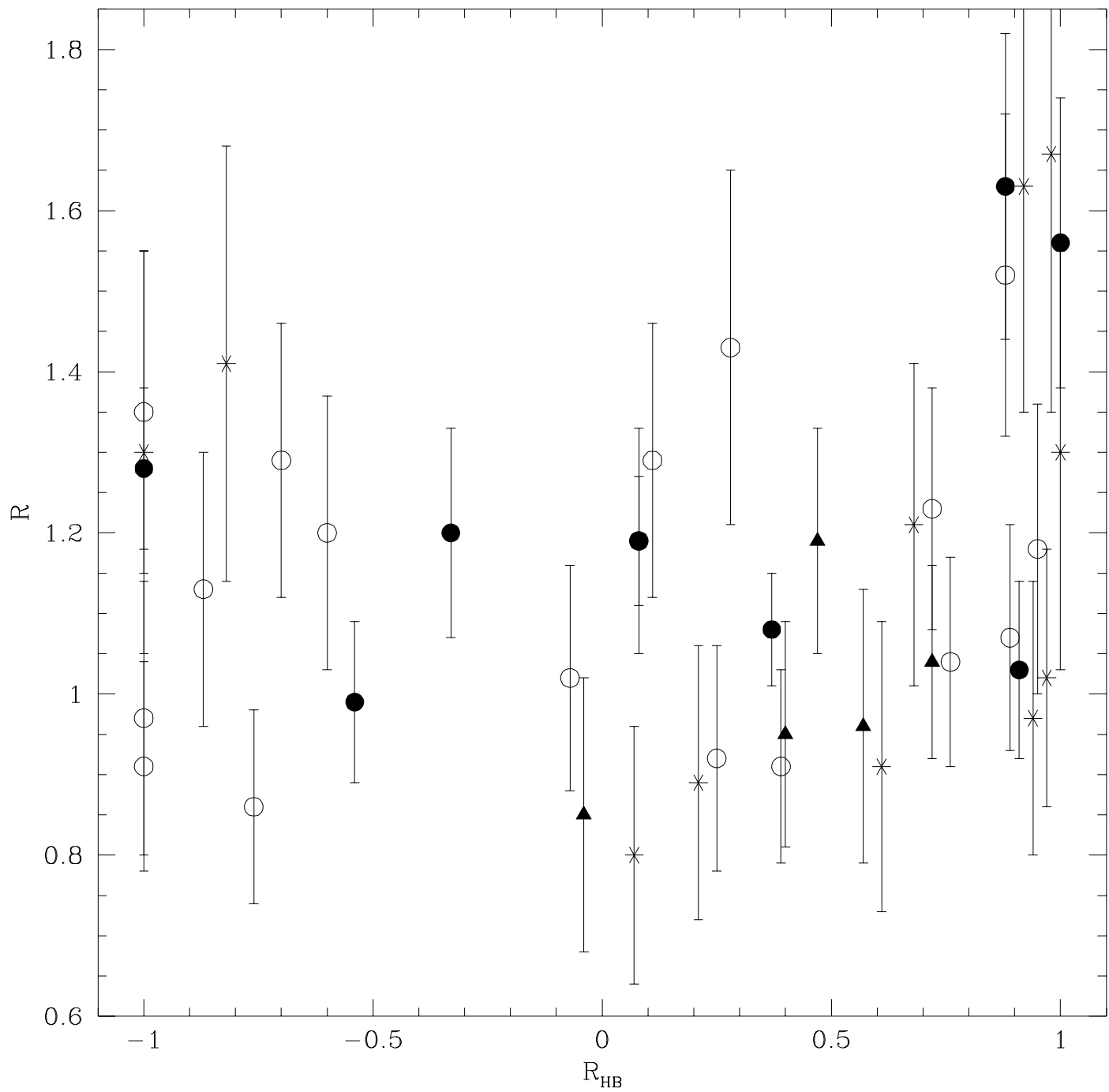
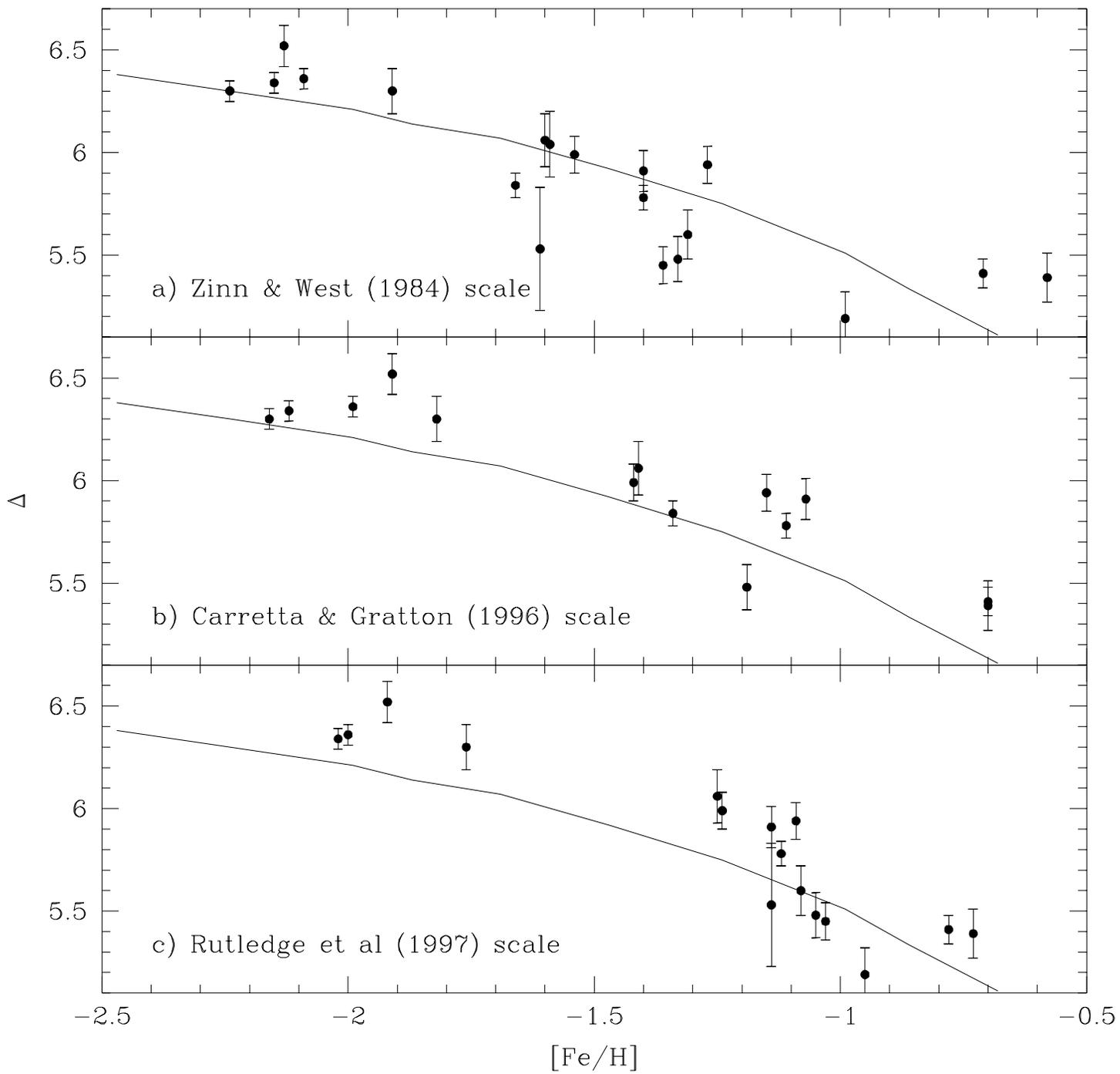
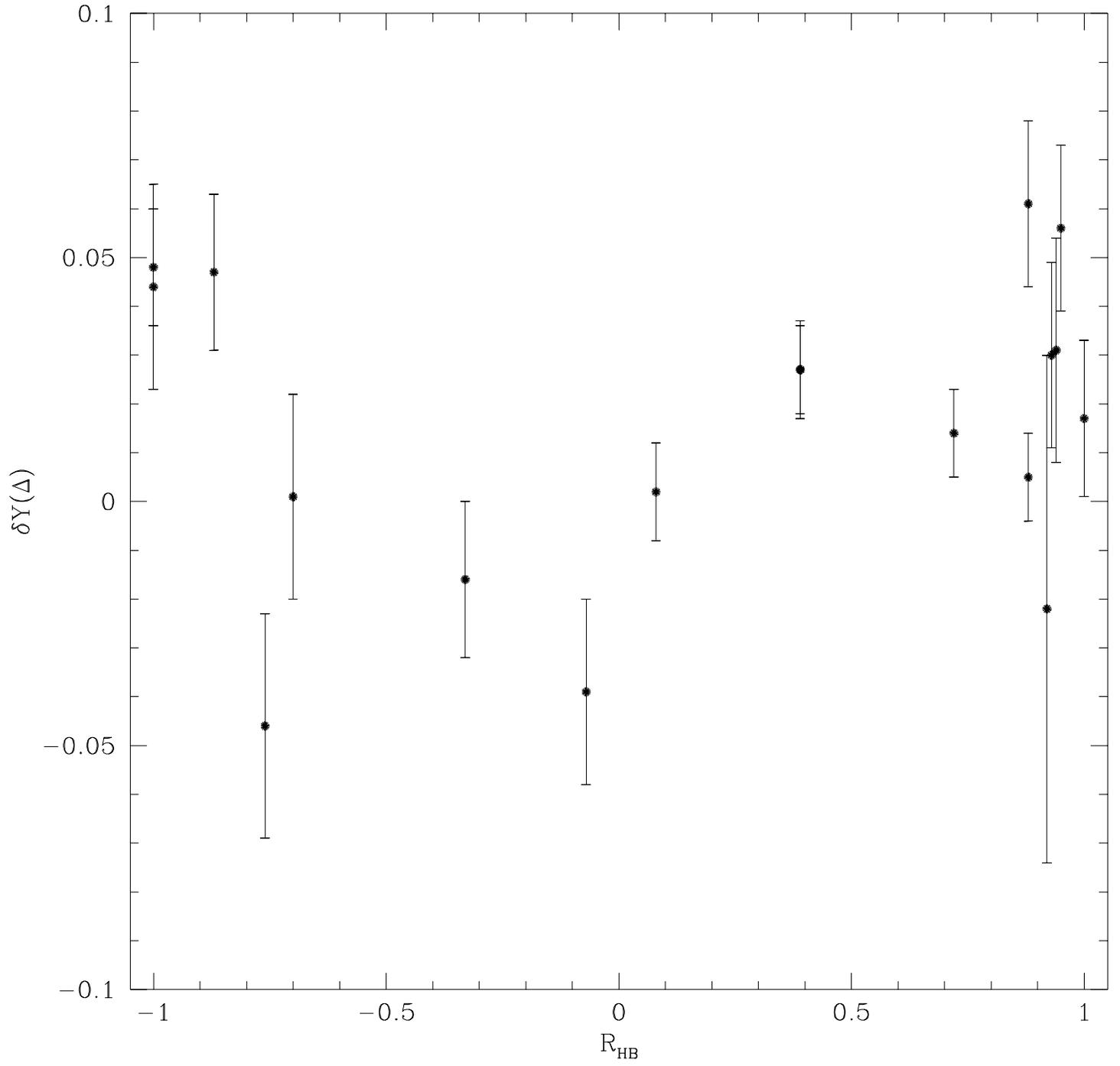
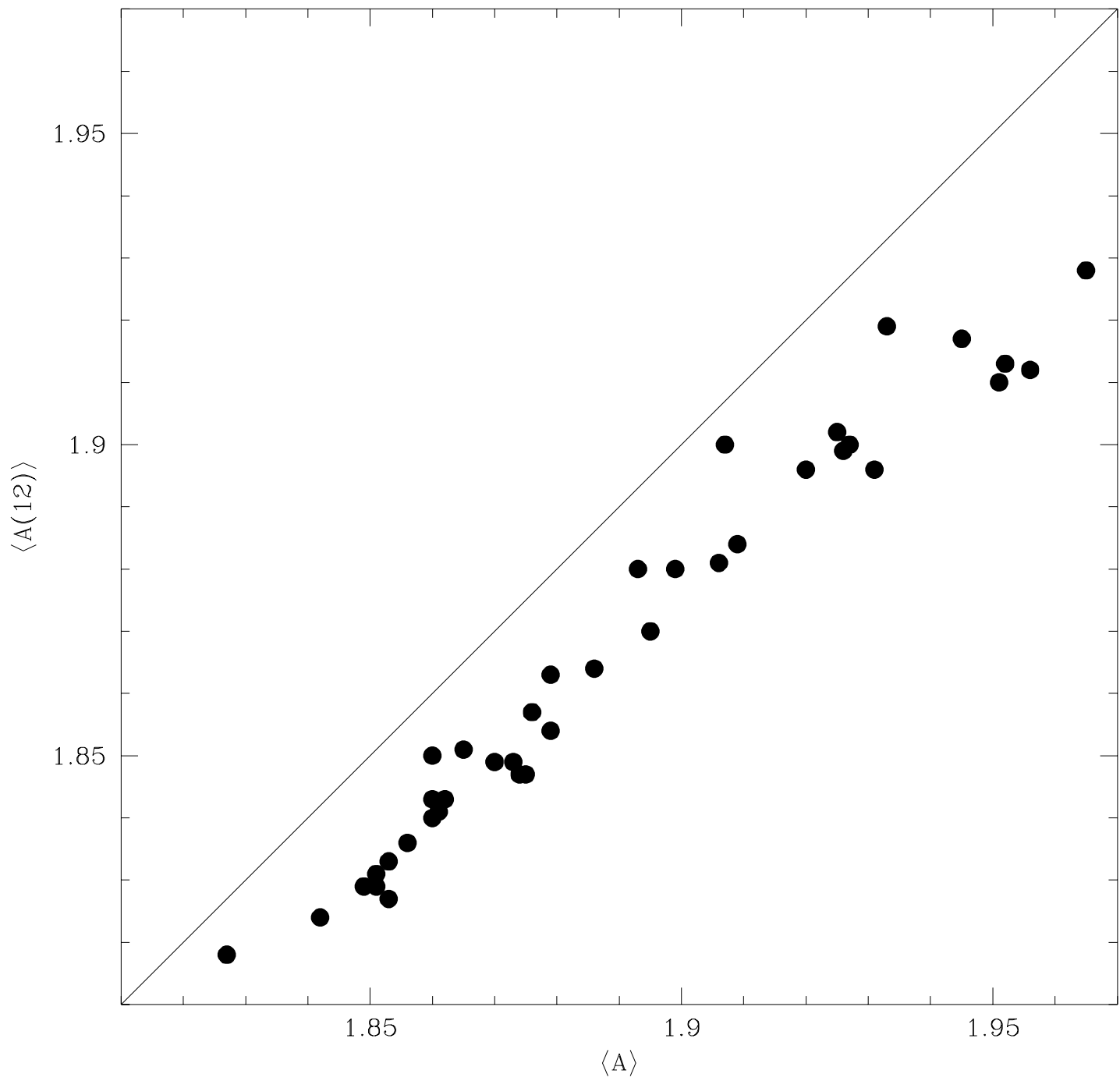
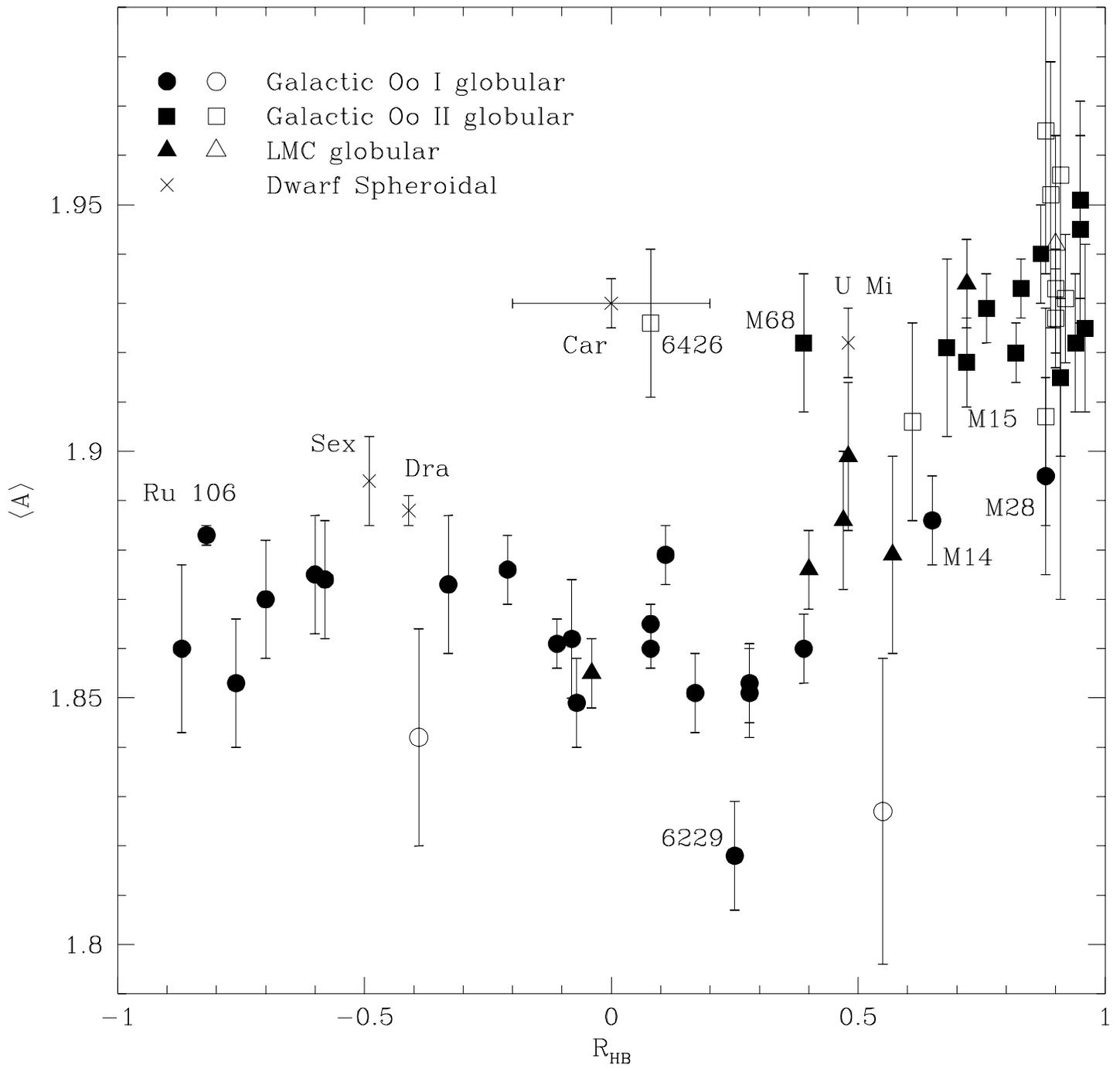
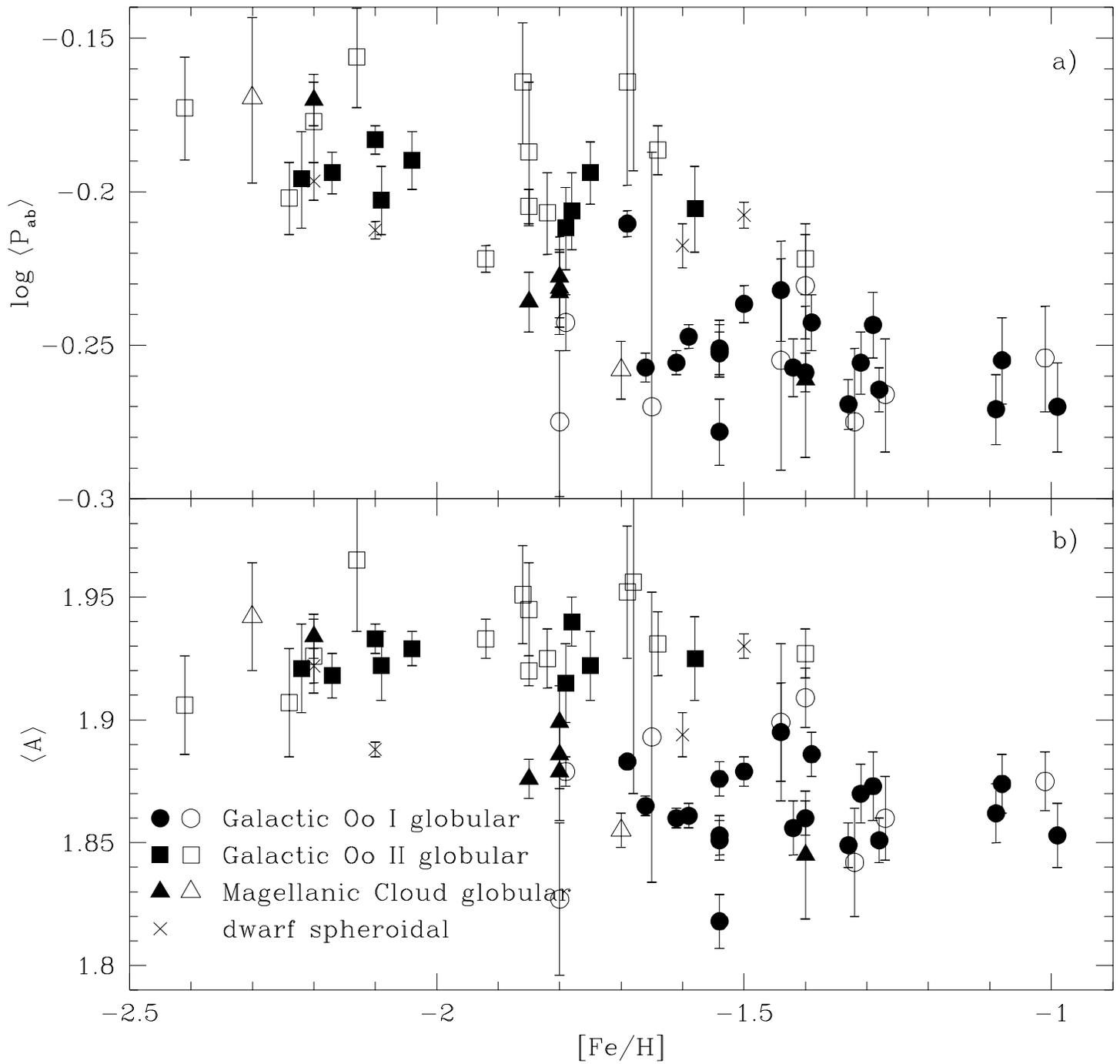
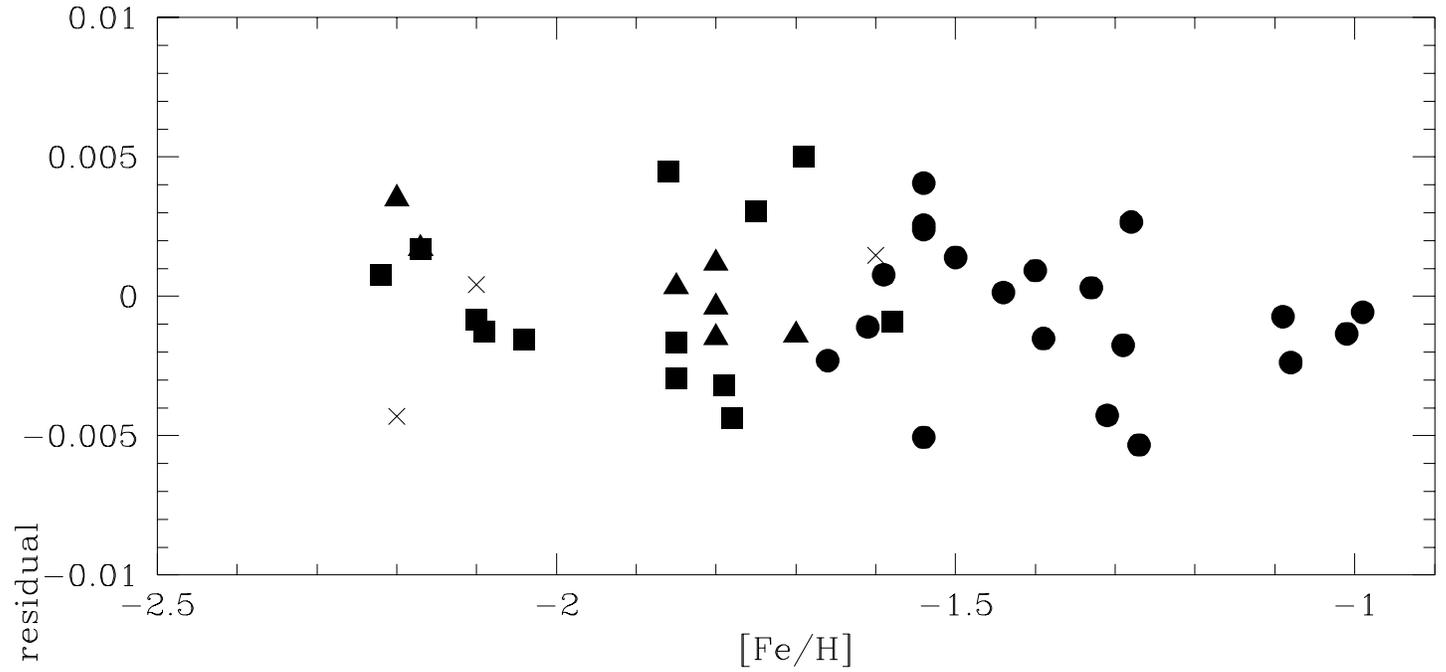
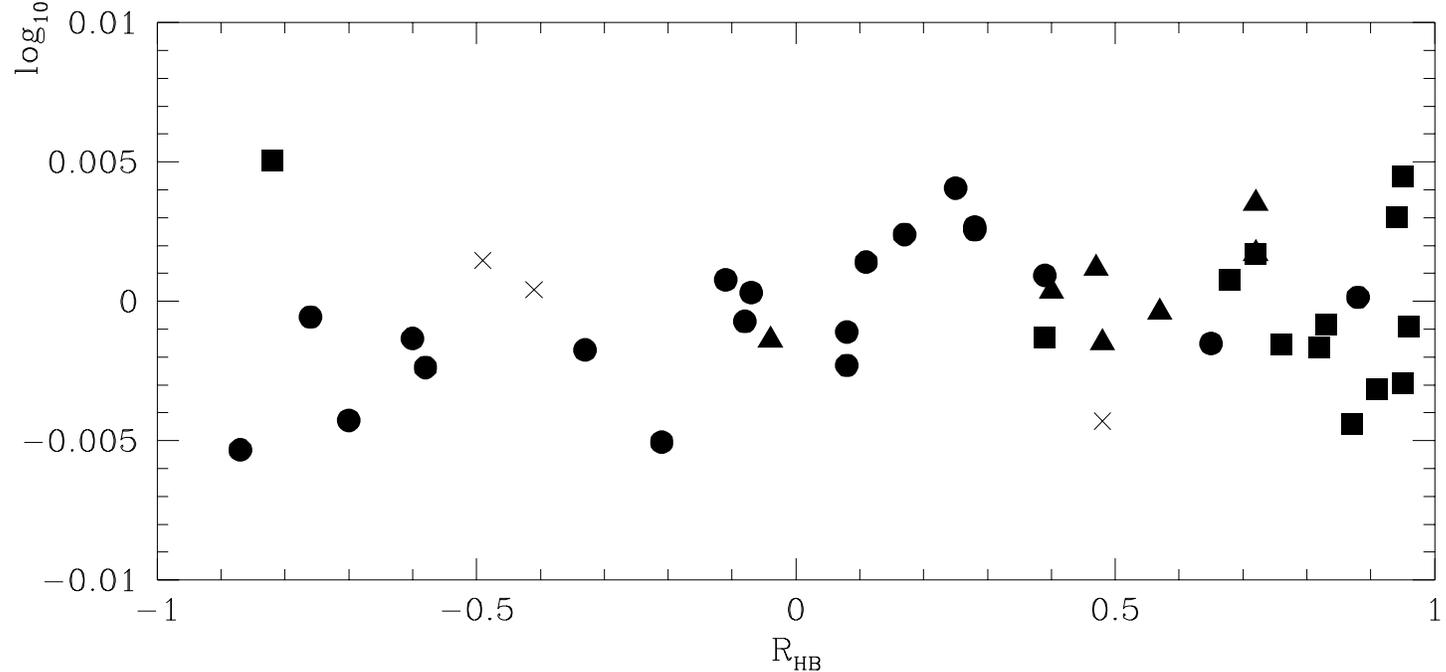
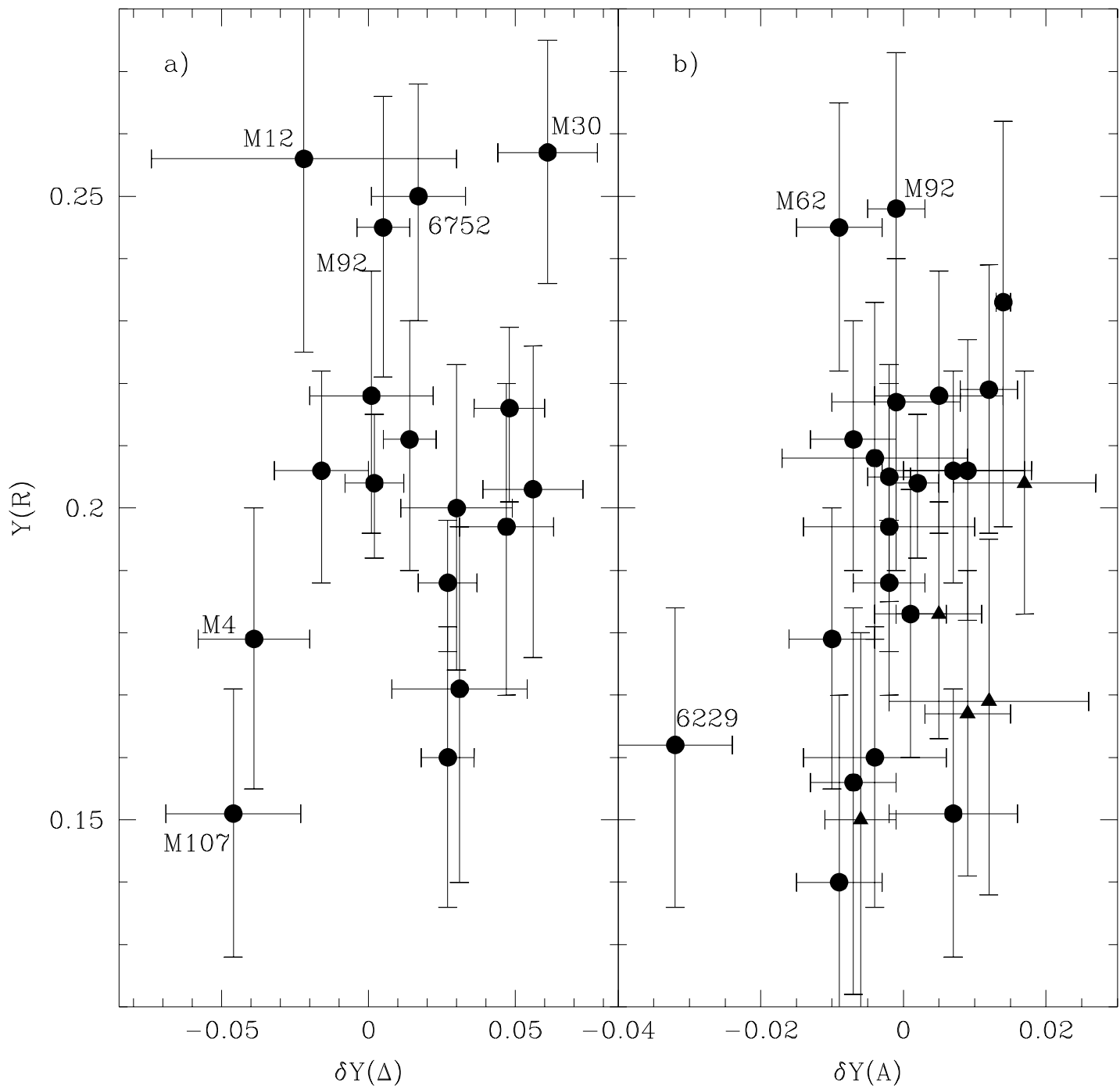
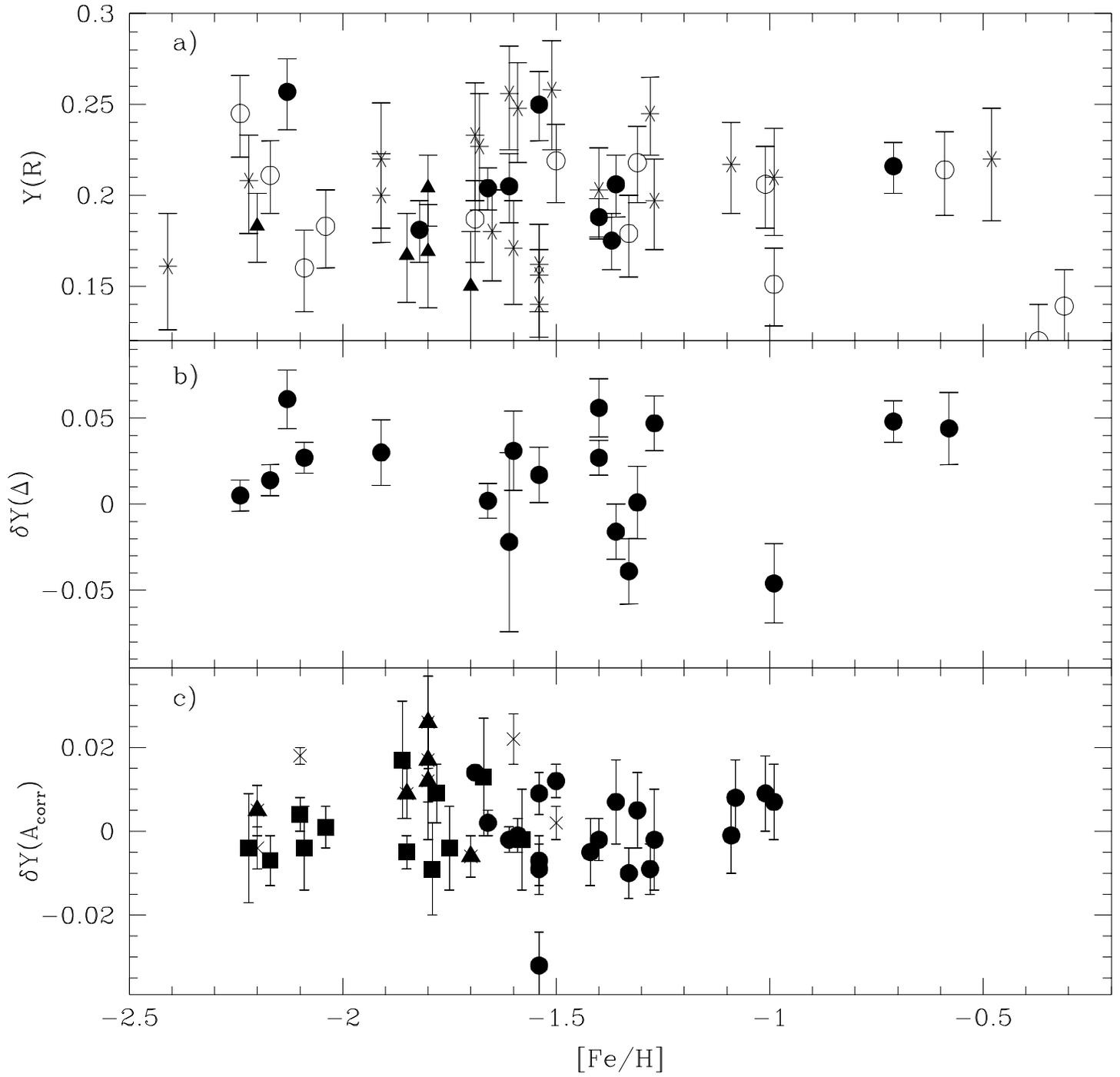
\end{document}